\documentclass{jpsj2}
\usepackage{bm}

\title{First-Principles Momentum Dependent Local Ansatz Approach to the
Momentum Distribution Function in Iron-Group Transition Metals}

\author{Yoshiro Kakehashi\thanks{yok@sci.u-ryukyu.ac.jp, to be published
in J. Phys. Soc. Jpn. (2017).} and Sumal Chandra}
\inst{Department of Physics and Earth Sciences, Faculty of Science,\\
University of the Ryukyus, \\
1 Senbaru, Nishihara, Okinawa, 903-0213, Japan} 

\abst{
The momentum distribution function (MDF) bands of iron-group transition
 metals from Sc to Cu have been investigated on the basis of the
 first-principles momentum dependent local ansatz wavefunction method. 
It is found that the MDF for $d$ electrons show a strong momentum
 dependence and a large deviation from the Fermi-Dirac distribution
 function along high-symmetry lines of the first Brillouin zone, 
while the $sp$ electrons
 behave as independent electrons.  In particular, the deviation in bcc
 Fe (fcc Ni) is shown to be enhanced by the narrow $e_{g}$ ($t_{2g}$)
 bands with flat dispersion in the vicinity of the Fermi level.
Mass enhancement factors (MEF) calculated from the jump on the Fermi
 surface are also shown to be momentum dependent.  Large
 mass enhancements of Mn and Fe are found to be caused by spin
 fluctuations due to $d$ electrons, while that for Ni is mainly caused
 by charge fluctuations.  Calculated MEF are consistent with
 electronic specific heat data as well as the recent angle resolved
 photoemission spectroscopy data.  
}

\kword{ first-principles variational theory, momentum-dependent local
ansatz, iron-group transition metals, electron correlations, momentum
distribution function, mass enhancement factor }

\begin{document}
\maketitle

\section{Introduction}

The iron-group transition metals and compounds show a variety of 
physical properties such as anomalous cohesive properties~\cite{jfcm77}, 
appearance of the ferro- and antiferro-magnetism~\cite{fulde12,kake13}, 
and high-$T_{\rm c}$ superconductivity~\cite{imada10}.
Many of their electronic, cohesive, and magnetic properties are 
well-known to be explained quantitatively by the density functional 
band theory (DFT)~\cite{hohen64,kohn65,rmm12,janak76,janak77,moruzzi78}.

The DFT is based on the Hohenberg-Kohn theorem which states that the
ground state is given by the functional of electron density and the
Kohn-Sham scheme which makes use of the density of an independent
electron system.
With use of the exchange-correlation potential in the local density
approximation (LDA)~\cite{barth72} or the generalized 
gradient approximation (GGA)~\cite{perdew86,bagno89}, 
the DFT quantitatively explained the stability of the structure and 
magnetism, the lattice parameter, the bulk modulus, as well as the
magnetism in transition metals and 
compounds~\cite{janak76,janak77,moruzzi78}.

Although the DFT has been successful in quantitative description of the
physical properties of many metals and compounds, problems and 
limitations of the DFT have also been clarified over the past five
decades.   
One of the serious problems is that the quantitative aspects of
the DFT become unstable with increasing Coulomb interaction strength. 
The DFT, for example, fails to explain the paramagnetism 
in $\epsilon$-Fe~\cite{pour14}, 
the weak antiferromagnetism in Fe-pnictides~\cite{imada10}, 
as well as the antiferromagnetism in cuprates~\cite{fulde12}.  
The problem limits the application range of the DFT.
The second problem is that excited states and related excitation 
spectra cannot be described by the DFT because the latter is 
based on the Hohenberg-Kohn theorem.
For the same reason, the physical quantities such as the 
charge and spin fluctuations described by the two-particle operators 
cannot be obtained by the DFT.
Finally, the momentum distribution function and
related mass enhancement factor cannot be obtained by the DFT because
the DFT is based on the Kohn-Sham scheme.

Because of the problems and limitations of the DFT mentioned above, 
the ground-state properties and related excitations 
of iron-group transition metals have not yet been fully understood 
from the quantitative point of view.
In order to clarify the properties, we have to take
alternative approaches such as the Gutzwiller wavefunction 
method~\cite{mcg63,mcg64,bune00,bune12,schick12,yk14} and  
the dynamical mean field theory (DMFT)~\cite{kotliar06,anisimov10}, 
or equivalently the dynamical coherent potential
approximation (DCPA)~\cite{kake08-1,kake10,kake11,kake13}.
Using the first-principles DCPA, we recently performed the calculations
of single-particle excitation spectra from Sc to Cu 
at finite temperatures, and elucidated the systematic 
change of the XPS spectra of iron-group transition metals~\cite{kake10}. 

For the quantitative description of the ground-state properties, the
wavefunction method is useful~\cite{fulde12,fulde16}.
The first-principles Gutzwiller theory can 
resolve a small energy difference between the states at zero temperature
which is not achieved by the first-principles DMFT.  
But it does not yield the correct weak Coulomb interaction limit. 
In order to describe quantitatively the ground-state properties of
correlated electrons, we have recently proposed the momentum 
dependent local ansatz (MLA) theory~\cite{kake08,pat11,pat13,yk14} 
which goes beyond the Gutzwiller wavefunction 
method~\cite{mcg63,mcg64}.
The MLA takes into account all the two-particle excited
states with momentum-dependent variational amplitudes, so that 
the theory reduces to the Rayleigh-Schr\"{o}dinger perturbation theory 
in the weak Coulomb interaction limit and describes well correlated 
electrons from the weak to strong Coulomb interaction regime.  
In particular, the MLA describes quantitatively the momentum
distribution function in contradiction to the case of the Gutzwiller
wavefunction.  

Quit recently, we extended the MLA to the first-principles version 
combining the theory with
the first-principles tight-binding LDA+U 
Hamiltonian~\cite{kake16,chan16}. 
On the basis of the first principles MLA, we calculated the
correlation energy, charge fluctuations, amplitude of local
moment, and the momentum distribution function for paramagnetic iron, 
and clarified the ground-state property~\cite{chan16}.
Subsequently, we investigated the correlation energy, charge
fluctuations, and the amplitude of local moment of the iron-group
transition metals in the paramagnetic state using the same theory, 
and clarified the correlation effects on these
quantities~\cite{kake16-2}. 

In this paper, we investigate the momentum distribution functions (MDF)
and mass enhancement factors (MEF) of the iron-group transition metals
from Sc to Cu on the basis of the first-principles MLA in order to
understand their systematic change over the $3d$ series.  
The MDF is the simplest static  quantity which cannot be 
described by the DFT and the simplest one-particle quantity 
indicating the strength of electron correlations. 
It also provides us with a Fermi liquid parameter of the system, 
{\it i.e.}, the MEF, from the jump at the Fermi surface.
Present work is the first systematic investigations 
for the change of the band structure of the MDF and the MEF
in iron-group transition metals at zero temperature.
We will demonstrate that the MDF bands for $d$ electrons in Mn, Fe, Co, 
and Ni strongly deviate from the Fermi-Dirac distribution function 
due to electron correlations.
These deviations yield significant MEF which cannot be explained by the
DFT. 

We remark that the first-principles MLA is competitive with the 
first-principles DMFT ({\it i.e.}, 
DCPA)~\cite{kotliar06,anisimov10,kake08-1,kake10,kake11,kake13} 
for the calculation of the properties at zero temperature. 
The DMFT is a powerful method to strongly correlated electrons and 
has been applied to many systems.  The accuracy of the DMFT however 
strongly depends on the solver of the impurity problem for correlated 
electrons.  The Quantum Monte-Carlo method (QMC) can describe accurately 
the finite-temperature properties of the system.  But its efficiency 
is strongly reduced at low temperatures, and the QMC even causes the 
negative sign problem which prevents us from systematic investigations 
over wide range of interaction parameters.  The exact diagonalization 
method (ED) is useful to study exactly the physical properties at zero 
temperature.  But it cannot describe the low energy properties 
associated with the Fermi surface.  The numerical renormalization group 
theory (NRG) can describe accurately the low energy excitations, but 
it does not accurately describe the excitations in high-energy region 
as well as the energy-integrated quantities.  Furthermore it is not 
applicable to the realistic systems because of the numerical difficulty.

The MLA describes quantitatively the 
quasi-particle weight associated with the low energy excitations 
as well as the energy-integrated quantities 
such as the total energy and momentum distribution function without
numerical difficulty.  
In particular, we have shown in the recent paper~\cite{kake16} 
that the first-principles MLA quantitatively explains 
the mass enhancement factor of bcc Fe obtained by the ARPES 
experiment, while the LDA+DMFT combined with the three-body theory at 
zero temperature does not~\cite{sanchez09}.  
Furthermore it also allows us to calculate 
any static physical quantity because the wavefunction is known.  
These facts indicate that the first-principles MLA is suitable for the
quantitative investigations of correlated electrons at zero temperature.

In the following section, we outline the first-principles MLA.  We 
present the MLA wavefunction with three kinds of correlators 
for the first-principles LDA+U Hamiltonian, and obtain the
ground-state energy in the single-site approximation (SSA).  Next, we
derive the self-consistent equations for variational parameters from 
the ground-state energy.  
In \S 3, we present the MDF calculated along high-symmetry lines 
in the first Brillouin zone.  
We demonstrate that the MDF for $d$ electrons show a large deviations
from the Fermi distribution function for Mn, Fe, Co, and Ni due to
electron correlations.  Accordingly, the MEF show significantly large
values from 1.2 to 1.7.
Calculated MEF are consistent with those
obtained from the electronic specific heat and angle resolved
photoemission spectroscopy (ARPES) data.
We will clarify the role of electron 
correlations in the MDF and MEF from Sc to Cu as well as
the role of $s$, $p$, and $d$ electrons in the MDF and the MEF.
In the last section we summarize our results and discuss the
effects of magnetism on the MEF.

\section{First-Principles MLA}

We adopt the first-principles LDA+U Hamiltonian with an atom 
in the unit cell~\cite{kake13,anisimov10}. 
\begin{align}
H & = \sum_{iL\sigma}\epsilon^{0}_{L} \ n_{iL\sigma} 
+ \sum_{iLjL^{'}\sigma}{t}_{iLjL^{'}}\ 
a^{\dagger}_{iL\sigma}\,{a}_{jL^{'}\sigma}  \nonumber \\
  & \hspace*{-4mm} + \sum_{i}\Big[ 
\sum_{m} {U}_{mm}  n_{ilm\uparrow}\, n_{ilm\downarrow} 
+ \! \sum_{(m,m')} \!\! \Big(U_{mm'}-\frac{1}{2}J_{mm'}\Big) 
n_{ilm} n_{ilm{'}} - 
 2 \!\! \sum_{(m,m')} \!\! J_{mm'}\,
\boldsymbol{s}_{ilm} \! \cdot \! \boldsymbol{s}_{ilm'} \Big] \,.
\label{eqhldau}
\end{align}
Here $\epsilon^{0}_{L}$ is an atomic level of orbital $ L$ on site $i$. 
${t}_{iLjL'}$  is a transfer integral between $iL$ and $jL'$,  
$L=(l, m)$ being the $s \,(l=0)$, $p\, (l=1)$, and $d \,(l=2)$ 
orbitals~\cite{ander84,ander85}. 
$a^{\dagger}_{iL\sigma}{({a}_{iL\sigma})}$ is the creation (annihilation) 
operator for an electron on site $i$ with orbital $L$ and spin 
${\sigma}$, and 
$n_{iL\sigma} = a^{\dagger}_{iL\sigma}{a}_{iL\sigma}$ is 
the number operator on the same site $i$ with orbital $L$ and 
spin $\sigma$.  
The atomic level $\epsilon^{0}_{L}$ is calculated from the LDA atomic 
level $\epsilon_{L}$ by subtracting the double counting 
potential~\cite{anisimov10}.
The third term at the rhs (right-hand-side) of Eq. (\ref{eqhldau})
denotes the on-site Coulomb interactions between $d$ electrons.
$U_{mm}\,(U_{mm'} )$ and $J_{mm'}$ are the intra-orbital (inter-orbital) 
Coulomb and exchange interactions between $d$ electrons, respectively. 
$n_{ilm}\, (\boldsymbol{s}_{ilm})$ with $l = 2 $ is the charge 
(spin) density operator for $d$ electrons on site $i$ and orbital $m$. 
The operator $\boldsymbol{s}_{{iL}}$ is defined as 
$\boldsymbol{s}_{iL} = \sum_{\gamma \gamma'} a^{\dagger}_{iL\gamma} 
(\boldsymbol{\sigma})_{\gamma\gamma'}\, {a}_{iL\gamma'}/2$, 
$\boldsymbol{\sigma}$ being the Pauli spin matrices.

In the first-principles MLA, we split the Hamiltonian $H$ into 
the Hartree-Fock part $H_{0}$ and the residual interaction part 
$H_{\mathrm {I}}$: 
\begin{equation}
H = H_{0} + H_{\mathrm{I}} \,.
\label{eqhhfi}
\end{equation}
The latter is expressed as follows.
\begin{align}
H_{\mathrm{I}} &= \sum_{i} \Big[ \sum_{L} U_{LL}^{(0)}\ {O}^{(0)}_{iLL} 
+ \sum_{(L,\, L')} U_{LL'}^{(1)} \ {O}^{(1)}_{iLL'} + 
\sum_{(L,\, L')} U_{LL'}^{(2)}\ {O}^{(2)}_{iLL'} \Big] \,.  
\label{eqhi}
\end{align}
The first term is the intra-orbital Coulomb interactions, 
the second term is the inter-orbital charge-charge interactions, 
and the third term denotes the inter-orbital spin-spin interactions, 
respectively. 
The Coulomb interaction energy parameters $U_{LL'}^{(\alpha)}$ are 
defined by $U_{LL}\delta_{LL'}$ $(\alpha=0)$, 
$U_{LL'}-J_{LL'}/2$ $(\alpha=1) $, 
and $-2J_{LL'}$ $ (\alpha=2)$, respectively. 
The operators ${O}^{(0)}_{iLL}$, ${O}^{(1)}_{iLL'}$, 
and ${O}^{(2)}_{iLL'}$  are defined by
\begin{equation}
{O}^{(\alpha)}_{iLL^{\prime}} = 
\begin{cases}
		 \ \delta n_{ilm\uparrow} \, 
\delta n_{ilm\downarrow} \, \delta_{LL^{\prime}} &  \ (\alpha=0) \\
\ \delta n_{ilm} \, \delta n_{ilm'}  &  \ (\alpha=1) \\ 
\ \delta \boldsymbol{s}_{ilm} \cdot 
\delta \boldsymbol{s}_{ilm'} & \ (\alpha=2)\, .
\end{cases}
\label{eqoalph}
\end{equation}
Note that $\delta A$ for an operator $A$ is defined by 
$\delta A = A-\langle A\rangle_{0}$, $\langle \sim \rangle_{0}$ being 
the average in the Hartree-Fock approximation. 

When the Hamiltonian $H$ is applied to the Hartree-Fock wavefunction
$|\phi \rangle$, the Hilbert space is expanded by the local operators 
$\{ O^{(\alpha)}_{iLL^{\prime}} \}$ in the interactions.
In order to take into account these states as well as the states
produced in the weak Coulomb interaction limit, we introduce the 
momentum-dependent local correlators 
$\lbrace\tilde{O}^{(\alpha)}_{iLL'}\rbrace$ ($\alpha=$ 0, 1, and 2) 
as follows.
\begin{align}
\tilde{O}^{(\alpha)}_{iLL'}&= \sum_{\{kn\sigma\}}\langle{k'_{2}n'_{2}\vert iL}\rangle_{\sigma'_{2}} \langle{iL\vert {k}_{2}{n}_{2}}\rangle_{\sigma_{2}} \langle{k'_{1}n'_{1}\vert iL'}\rangle_{\sigma'_{1}} \langle{iL'\vert {k}_{1}{n}_{1}}\rangle_{\sigma_{1}}  \nonumber \\
&\hspace{1cm}\times\lambda^{(\alpha)}_{{LL'}\{{2'2 1'1}\}}\ \delta(a^{\dagger}_{k'_{2}n'_{2}\sigma'_{2}}a_{{k}_{2}{n}_{2}\sigma_{2}})\ \delta(a^{\dagger}_{k'_{1}n'_{1}\sigma'_{1}}a_{{k}_{1}{n}_{1}\sigma_{1}})\,.
\label{eqotilde}
\end{align}
Here $a^{\dagger}_{k n \sigma }{(a_{k n\sigma})}$ is the creation 
(annihilation) operator for an electron with momentum $\bm{k}$, 
band index $n$, and spin $\sigma $. These operators are given by those 
in the site representation as 
$a_{k n\sigma}=\sum_{iL}a_{iL\sigma}\langle k n\vert iL\rangle_{\sigma}$\,. 
$\langle k n\vert iL\rangle_{\sigma}$ are the overlap integrals
between the Bloch state $(\bm{k}n)$ and the local-orbital state $(iL)$.

The momentum-dependent parameters 
$\lambda^{(\alpha)}_{{LL'}\{{2'2 1'1}\}}$ in Eq. (\ref{eqotilde}) are
defined as
\begin{equation}
\lambda^{(0)}_{{LL'}\{{2'2 1'1}\}} = 
\eta_{L [2'2 1'1]}
\ \delta_{LL'}\,\delta_{\sigma'_{2}\downarrow}
\,\delta_{\sigma_{2}\downarrow}\,\delta_{\sigma'_{1}\uparrow}
\,\delta_{\sigma_{1}\uparrow}\,,
\label{eqlam0}
\end{equation}
\begin{equation}
\hspace{-1.5cm}\lambda^{(1)}_{{LL'}\{2'2 1'1\}} = 
\zeta^{(\sigma_{2}\sigma_{1})}_{LL' [2'2 1'1]}
\ \delta_{\sigma'_{2}\sigma_{2}}\, \delta_{\sigma'_{1}\sigma_{1}}\,,
\label{eqlam1}
\end{equation}
\begin{align}
\lambda^{(2)}_{{LL'}\{{2'2 1'1}\}} & = 
\sum_{\sigma}\xi^{(\sigma)}_{LL' [2'2 1'1]}
\ \delta_{\sigma'_{2}-\sigma}\  \delta_{\sigma_{2}\sigma}
\ \delta_{\sigma'_{1}\sigma}\ \delta_{\sigma_{1}-\sigma} \nonumber \\
&\hspace{1cm} + \frac{1}{2} \sigma_{1} \sigma_{2} 
\ \xi^{(\sigma_{2}\sigma_{1})}_{LL' [2'2 1'1]}
\ \delta_{\sigma'_{2}\sigma_{2}} \,\delta_{\sigma'_{1}\sigma_{1}}\,.
\label{eqlam2}
\end{align}
Here $\{2'21'1\}$ ($[2'21'1]$) implies that 
$\{2'21'1\} = k'_{2}n'_{2}\sigma'_{2}k_{2}n_{2}\sigma_{2} 
k'_{1}n'_{1}\sigma'_{1}k_{1}n_{1}\sigma_{1}$
( $[2'21'1] = k'_{2}n'_{2}k_{2}n_{2}k'_{1}n'_{1}k_{1}n_{1}$). 
$\eta_{L [2'21'1]}$, $\zeta^{(\sigma_{2}\sigma_{1})}_{L L'[2'21'1]}$, 
$\xi^{(\sigma)}_{L L'[2'21'1]}$, and 
$\xi^{(\sigma_{2}\sigma_{1})}_{L L'[2'21'1]}$ are the variational
parameters.
Note that 
$\tilde{O}^{(0)}_{iLL}$, $\tilde{O}^{(1)}_{iLL'}$, and 
$\tilde{O}^{(2)}_{iLL'}$ reduce to the local correlators, 
${O}^{(0)}_{iLL}$, ${O}^{(1)}_{iLL'}$, and ${O}^{(2)}_{iLL'}$ when 
$\eta_{L [2'21'1]} = \zeta^{(\sigma_{2}\sigma_{1})}_{L L'[2'21'1]}=1$ and 
$\xi^{(\sigma)}_{L L'[2'21'1]} = 
\xi^{(\sigma_{2}\sigma_{1})}_{L L'[2'21'1]}=1/2$, so that 
$\{ \tilde{O}^{(\alpha)}_{iLL'} \}$ describe the intra-orbital 
correlations, 
the inter-orbital charge-charge correlations, and the inter-orbital
spin-spin correlations (, $i.e.,$ the Hund-rule correlations),
respectively.

Using the correlators $\lbrace\tilde{O}^{(\alpha)}_{iLL'}\rbrace$ and 
the Hartree-Fock ground-state wavefunction $\vert{\phi}\rangle$, 
we construct the first-principles MLA wavefunction as follows.
\begin{equation}
\vert{\Psi}_\mathrm{MLA}\rangle = {\Big[\prod_{i}{ \Big( 
1 - \sum_{L}{\tilde{O}}^{(0)}_{iLL} - 
\sum_{(L,L')}{\tilde{O}}^{(1)}_{iLL'} - 
\sum_{(L,L')}{\tilde{O}}^{(2)}_{iLL'} \Big) } \Big]} 
\ \vert{\phi} \rangle \,.
\label{eqmla}
\end{equation}
We note that the MLA wavefunction reduces to the local ansatz
(LA) wavefunction by Stollhoff and 
Fulde~\cite{gs77,gs78,gs80,stoll81,oles84,capell86}, 
when the variational
parameters $\lambda^{(\alpha)}_{{LL'}\{{2'2 1'1}\}}$ are momentum
independent. The momentum dependence of the variational parameters is 
taken into account in order to describe exactly the weak Coulomb 
interaction limit. 

The ground-state energy $\langle H \rangle$ is given by

\begin{align}
\langle H \rangle = \langle H\rangle_{0} + N\epsilon_c \,.
\label{eqener}
\end{align}
Here $\langle H \rangle_{0}$ denotes the Hartree-Fock energy, $N$ is the
number of atoms in the system.  $\epsilon_c$ is the correlation energy 
per atom defined by 
$N\epsilon_c \equiv \langle \tilde{H} \rangle = 
\langle H \rangle-\langle H \rangle_{0}$. 
Note that 
$\tilde{H}\equiv H -\langle H \rangle_{0} = \tilde{H}_{0}+H_{\rm I}$. 
$\langle \sim \rangle$ ($\langle \sim \rangle_{0}$) denotes the full 
(Hartree-Fock) average with respect to 
$\vert \Psi_\mathrm{MLA}\rangle$ ($\vert \phi \rangle$).
The correlation energy $\epsilon_c$ is expressed in the single-site 
approximation (SSA) as follows~\cite{chan16,yk14}. 
\begin{equation}
{\epsilon_c} = \frac{{- \langle {\tilde{O_i}^\dagger}} {H}_{\rm I}\rangle_0 
- \langle {H}_{\rm I} \tilde{O_i}\rangle_0 
+ \langle {\tilde{O_i}^\dagger} \tilde{H} \tilde{O_i} \rangle_0 }
{1 + \langle \tilde{O_i}^\dagger\tilde{O_i} \rangle_0} \,. 
\label{eqcorr}
\end{equation}
Here $\tilde{O_i} = \sum_{\alpha} \sum_{\langle L,\, L' \rangle} 
\tilde{O}^{(\alpha)}_{iLL'}$. 
The sum $\sum_{\langle L,\, L' \rangle}$ is defined by a single sum 
$\sum_{L}$ when  $L'$=$L$, and by a pair sum $\sum_{(L,\, L')}$ 
when $L' \neq L$.
Each element in Eq. (\ref{eqcorr}) has been calculated with use of 
Wick's theorem~\cite{chan16}.

The variational parameters are determined from the stationary condition 
$\delta\epsilon_{c}=0$ as follows.
\begin{equation}
-\langle(\delta \tilde{O}_{i}^{\dagger}){H}_{\rm I}\rangle_0 + 
\langle (\delta\tilde{O}_{i}^{\dagger} )\tilde{H} \tilde{O_i}\rangle_0 -
\epsilon_c {\langle(\delta\tilde{O}_{i}^{\dagger}) \tilde{O_i}\rangle_0}
+ {\rm c.c.} = 0 \,.
\label{eqvar}
\end{equation}
Here $\delta \tilde{O}_{i}^{\dagger}$ denotes the variation of 
$\tilde{O}_{i}^{\dagger}$ with respect to 
$\{ \lambda^{(\alpha)}_{{LL'}\{2'2 1'1\}} \}$.

Since it is not easy to solve Eq. (\ref{eqvar}) for arbitrary Coulomb
interaction strength, we
make use of the following ansatz for the variational parameters, 
which interpolates between the weak Coulomb interaction limit and 
the atomic limit~\cite{kake16,chan16}. 
\begin{equation}
\lambda^{(\alpha)}_{{LL'}\{{2'2 1'1}\}} = 
\frac{ U_{LL'}^{(\alpha)} \sum_{\tau} 
C_{\tau\sigma_{2}\sigma_{2}^{'}\sigma_{1}\sigma_{1}^{'}}^{(\alpha)}
\ \tilde{\lambda}_{\alpha\tau L L'}^{(\sigma_{2}\sigma_{1})} }
{ \epsilon_{k^{\prime}_{2} n^{\prime}_{2} \sigma^{\prime}_{2}} - 
\epsilon_{k_{2} n_{2} \sigma_{2}} + 
\epsilon_{k^{\prime}_{1} n^{\prime}_{1} \sigma^{\prime}_{1}} - 
\epsilon_{k_{1} n_{1} \sigma_{1}}
- \epsilon_{\rm c} } \,.
\label{eqvlam}
\end{equation}
Here the spin-dependent coefficients 
$C_{\tau\sigma_{2}\sigma_{2}^{'}\sigma_{1}\sigma_{1}^{'}}^{(\alpha)}$ 
are defined by 
$\delta_{\sigma'_{2}\downarrow}\,\delta_{\sigma_{2}\downarrow}\,
\delta_{\sigma'_{1}\uparrow}\,\delta_{\sigma_{1}\uparrow}$  
($\alpha=0$), 
$\delta_{\sigma'_{2}\sigma_{2}} \,\delta_{\sigma'_{1}\sigma_{1}}$  
($\alpha=1$), 
$-(1/4)\ \ \sigma_{1}\sigma_{2}\delta_{\sigma'_{2}\sigma_{2}} 
\delta_{\sigma'_{1}\sigma_{1}}$ ($\alpha=2,\tau=l$), and 
$-(1/2)\sum_{\sigma}\delta_{\sigma'_{2} - \sigma} 
\delta_{\sigma_{2}\sigma} \delta_{\sigma'_{1}\sigma} 
\delta_{\sigma_{1}-\sigma}$ ($\alpha=2,\tau=t$), respectively. 
Note that $l\,(t)$ implies the longitudinal (transverse) component. 
The renormalization factors 
$\tilde{\lambda}_{\alpha\tau L L'}^{(\sigma\sigma')}$ 
in Eq. (\ref{eqvlam}) are defined as 
$\tilde{\eta}_{LL'} \delta_{LL'}\delta_{\sigma'-\sigma}$ $(\alpha=0)$,  
$\tilde{\zeta}_{LL'}^{(\sigma\sigma')}$ $(\alpha=1)$, 
$\tilde{\xi}_{tLL'}^{(\sigma)}\delta_{\sigma'-\sigma}$ 
$(\alpha=2, \tau=t)$, and 
$\tilde{\xi}_{lLL'}^{(\sigma\sigma')}$ $(\alpha=2,\tau=l)$, 
respectively. 
The denominator in Eq. (\ref{eqvlam}) expresses the two-particle
excitation energy.  $\epsilon_{kn\sigma}$ denotes the Hartree-Fock one
electron energy eigenvalue for the momentum $\boldsymbol{k}$, 
the band index $n$, and spin $\sigma$.
Note that when 
$\tilde{\eta}_{LL} = \tilde\zeta^{(\sigma\sigma')}_{LL'} = 1$ 
and $\tilde \xi^{(\sigma\sigma')}_{lLL'} = 
\tilde \xi^{(\sigma)}_{tLL'} = -1$, the MLA wavefunction (\ref{eqmla})
reduces to that of the Rayleigh-Schr\"{o}dinger perturbation theory in
the weak Coulomb interaction limit.
The renormalization factors 
$\tilde{\eta}_{LL}$, $\tilde\zeta^{(\sigma\sigma')}_{LL'}$, 
$\tilde \xi^{(\sigma)}_{tLL'}$, and 
$\tilde \xi^{(\sigma\sigma')}_{lLL'}$ are the new variational 
parameters to be determined.

Substituting Eq. (\ref{eqvlam}) into the elements in Eq. (\ref{eqvar}), 
we obtain the self-consistent equations for the variational parameters.
In the paramagnetic case, the variational parameters 
$\tilde{\lambda}_{\alpha\tau L L'}^{(\sigma\sigma')}$ are spin 
independent (, $i.e., \tilde{\lambda}_{\alpha\tau L L'} $), 
and the self-consistent equations are expressed as 
follows~\cite{chan16}. 
\begin{equation}
\tilde{\lambda}_{\alpha\tau LL'} = 
\tilde{Q}_{LL'}^{-1} \left( \kappa_{\alpha} P_{LL'} - 
{U}^{(\alpha)\,-1}_{LL'} \, K^{(\alpha)}_{\tau LL'} \right) \, .
\label{eqsclam}
\end{equation}
Here $\tilde{Q}_{LL'}$ has the form 
$\tilde{Q}_{LL'}={Q}_{LL'}-{\epsilon_c} S_{LL'}$. 
The constant $\kappa_{\alpha}$ is defined by $1$ for $\alpha=0, 1$, and 
$-1$ for $\alpha=2$.
The second terms at the rhs of Eq. (\ref{eqsclam}) originates in the 
matrix element 
$\langle {\tilde{O_i}^\dagger} H_{\rm I} \tilde{O_i} \rangle_0$ 
in the numerator of the correlation energy (\ref{eqcorr}).
These terms are of higher order in Coulomb interactions and are given
by a linear combination of $\{ \tilde{\lambda}_{\alpha\tau LL'} \}$.
${Q}_{LL'}$, $S_{LL'}$, $P_{LL'}$, and $K^{(\alpha)}_{\tau LL'}$ are 
expressed by the Laplace transforms of the Hartree-Fock local densities 
of states~\cite{chan16}.

It should be noted that $\tilde{Q}_{LL'}$, $P_{LL'}$, and 
$K^{(\alpha)}_{\tau LL'}$ contain the correlation energy 
$\epsilon_{c}$ and the Fermi level $\epsilon_{\rm F}$.  
Moreover $K^{(\alpha)}_{\tau LL'}$ are given by a 
linear combination of $\{ \tilde{\lambda}_{\alpha\tau LL'} \}$.
The correlation energy $\epsilon_{c}$ is expressed by 
Eq. (\ref{eqcorr}) with variational parameters (\ref{eqvlam}). 
The Fermi level $\epsilon_{\rm F}$ is determined by the 
conduction electron number per atom $n_{e}$, which is expressed as
\begin{equation}
n_{e}=\sum_{L}\langle n_{iL}\rangle\,.
\label{eqne}
\end{equation}
Taking the same steps as in Eq. (\ref{eqcorr}), we obtain the 
partial electron number of orbital $L$ on site $i$ as follows in the SSA. 
\begin{equation}
\langle n_{iL}\rangle = \langle n_{iL}\rangle_{0} + 
\langle \tilde{n}_{iL}\rangle \,.
\label{eqnil}
\end{equation}
Here $\langle n_{iL}\rangle_{0}$ denotes the  Hartree-Fock electron 
number. The correlation correction $\langle \tilde{n}_{iL}\rangle$ is 
expressed as follows. 
\begin{align}
\langle \tilde{n}_{iL}\rangle = \frac{ \langle 
\tilde{O}_{i}^{\dagger} \tilde{n}_{iL} \tilde{O}_{i} \rangle_{0}}
{1 + \langle \tilde{O_i}^{\dagger}\tilde{O_i} \rangle_0 } \, .
\label{eqnilcorr}
\end{align}
Note that $\langle \tilde{O}_{i}^{\dagger} \tilde{n}_{iL} \rangle_{0}$
and $\langle \tilde{n}_{iL} \tilde{O}_{i}^{\dagger} \rangle_{0}$, 
which correspond to the first and second terms in the numerator of the
correlation energy (\ref{eqcorr}), vanish according to Wick's theorem.
The other elements at the rhs of Eq. (\ref{eqnilcorr}) 
are also calculated by using Wick's theorem.
Equations (\ref{eqcorr}), (\ref{eqsclam}), and (\ref{eqne}) determine 
self-consistently the correlation energy $\epsilon_{c}$, 
the Fermi level $\epsilon_{\rm F}$, 
as well as the variational parameters 
$\{\tilde{\lambda}_{\alpha\tau LL'}\}$. 

The momentum distribution function (MDF) is given as follows.
\begin{equation}
\langle n_{kn\sigma}\rangle = f(\tilde{\epsilon}_{kn\sigma}) 
+\frac{N\langle\tilde{O}_{i}^{\dagger} \tilde{n}_{kn\sigma} 
\tilde{O}_{i}\rangle_{0}}
{1+\langle{\tilde{O_i}^\dagger\tilde{O_i}}\rangle_0}\,.
\label{eqnkn}
\end{equation}
The first term at the rhs is the MDF for the Hartree-Fock independent 
electrons, {\it i.e.}, the Fermi distribution function (FDF) at 
zero temperature. 
$\tilde{\epsilon}_{kn\sigma}$ is the Hartree-Fock one-electron energy 
measured from the Fermi level $\epsilon_{\rm F}$. The second term 
is the correlation corrections, where $\tilde{n}_{kn\sigma}$ is defined 
by $\tilde{n}_{kn\sigma}= n_{kn\sigma}-\langle
n_{kn\sigma}\rangle_{0}$. 
The numerator has the following form~\cite{chan16}.
\begin{equation}
N \langle \tilde{O}_{i}^{\dagger} \tilde{n}_{kn\sigma} 
\tilde{O}_{i} \rangle_{0}
 = \sum_{\alpha\tau \ \langle L, L' \rangle} 
q_{\tau}^{(\alpha)}\ U_{LL'}^{(\alpha)2}\ 
\tilde{\lambda}_{\alpha\tau LL'}^{2}\ 
\big[ \hat{B}_{LL'n\sigma} (\boldsymbol{k})\ f(-\tilde{\epsilon}_{kn\sigma}) - 
\hat{C}_{LL'n\sigma} (\boldsymbol{k})\ f(\tilde{\epsilon}_{kn\sigma}) \big]\,.
\label{equ46a}
\end{equation}
Here $q_{\tau}^{(\alpha)}$ is a constant factor taking the value 1 for
$\alpha$=$0$, 2 for $\alpha$=$1$, 1/8 for $\alpha$=$2$, $\tau$=$l$, and
1/4 for $\alpha$=$2$, $\tau$=$t$, respectively. $\hat{B}_{LL'n\sigma}
(\boldsymbol{k})$ is a momentum-dependent particle contribution above 
$\epsilon_{\rm F}$ and is expressed as follows. 
\begin{equation}
\hat{B}_{LL'n\sigma} (\boldsymbol{k})=\vert u_{Ln\sigma}(\boldsymbol{k})\vert^{2}{B}_{L'L\sigma}({\epsilon}_{kn\sigma})+ \vert u_{L'n\sigma}(\boldsymbol{k})\vert^{2}{B}_{LL'\sigma}({\epsilon}_{kn\sigma})\,,
\label{eqnbk}
\end{equation}
where $\{u_{Ln\sigma}(\boldsymbol{k})\}$ are the eigenvectors for a
given $\boldsymbol{k}$ point. The hole contribution 
$\hat{C}_{LL'n\sigma} (\boldsymbol{k})$ is defined by 
Eq. (\ref{eqnbk}) in which the energy dependent terms 
${B}_{LL'\sigma} ({\epsilon}_{kn\sigma})$ have been replaced by 
${C}_{LL'\sigma} ({\epsilon}_{kn\sigma})$. These are given by the 
Laplace transformation of the local density of states in the 
Hartree-Fock approximation~\cite{chan16}. 
Note that the correlation correction to 
$\langle \tilde{n}_{kn\sigma}\rangle$ 
depends on $\boldsymbol{k}$ via
both energy $\tilde{\epsilon}_{kn\sigma}$ and eigenvector 
$u_{Ln\sigma}(\boldsymbol{k})$.

The quasiparticle weight $Z_{{k_{\rm F}}n}$ characterizes the low-energy 
excitations in metals. It is obtained by taking the difference between 
$\langle n_{kn\sigma}\rangle$ below and above the Fermi level 
$\epsilon_{\rm F}$. Taking average over the Fermi surface, we obtain 
the average quasiparticle weight $Z$.
\begin{equation}
Z = 1 + \frac{\overline{\delta(N \langle \tilde{O}_{i}^{\dagger} 
\tilde{n}_{kn\sigma}\tilde{O}_{i} \rangle_{0})_{k_{\rm F}}}}
{1+\langle{\tilde{O_i}^\dagger\tilde{O_i}}\rangle_0}\,.
\label{eqnavz}
\end{equation}
Here the first term at the rhs denotes the Hartree-Fock part. 
The second term is the correlation corrections. The upper bar in the 
numerator denotes the average over the Fermi surface, and 
${\delta(N\langle\tilde{O}_{i}^{\dagger}\tilde{n}_{kn\sigma} 
\tilde{O}_{i}\rangle_{0})_{k_{\rm F}}}$ 
means the amount of jump at the wavevector $\boldsymbol{k}_{\rm F}$ 
on the Fermi surface. 

In order to clarify the role of $s$, $p$, and $d$ electrons, it is
convenient to define the projected MDF for orbital $L$ by 
$\langle n_{kL\sigma}\rangle=\sum_{n}\langle n_{kn\sigma}\rangle\vert
u_{Ln\sigma}(\bm{k})\vert^{2}$. 
Furthermore, we replace the energy $\epsilon_{kn\sigma}$ in the 
expression with 
$\epsilon_{kL\sigma}=\sum_{n}\epsilon_{kn\sigma}\vert
u_{Ln\sigma}(\bm{k})\vert^{2}$, 
$i.e.,$ a common energy band projected onto the orbital $L$. We have then  
\begin{equation}
\langle n_{kL\sigma} \rangle = f(\tilde{\epsilon}_{kL\sigma}) 
+ \frac{N\langle\tilde{O}_{i}^{\dagger}\tilde{n}_{kL\sigma} 
\tilde{O}_{i}\rangle_{0}}
{1+\langle{\tilde{O_i}^\dagger\tilde{O_i}}\rangle_0} \,.
\label{eqnnkpro}
\end{equation}
We can also define the partial MDF $\langle n_{kl\sigma}\rangle$ for $l$
($= s,p,d$) electrons by
\begin{equation}
\langle n_{kl\sigma}\rangle = 
\frac{1}{2l+1} \sum_{m} \langle n_{kL\sigma} \rangle \,.
\label{eqnnkp}
\end{equation}
It should be noted that the projected MDF depend on the momentum
$\boldsymbol{k}$ only via $\tilde{\epsilon}_{kL\sigma}$. 

We can define the quasiparticle weight $Z_{L}$ for electrons with 
orbital symmetry $L$ by the jump of $\langle n_{kL\sigma}\rangle$ on 
the Fermi surface. 
\begin{align}
Z_{L}=1 + 
\frac{\overline{\delta(N\langle \tilde{O}_{i}^{\dagger} 
\tilde{n}_{kL\sigma} \tilde{O}_{i} \rangle_{0})}_{k_{\rm F}}}
{1+\langle{\tilde{O_i}^\dagger\tilde{O_i}}\rangle_0}\,.
\label{equ48}
\end{align}
Then we can verify the sum rule, 
\begin{align}
Z = \frac{1}{D}\sum_{L}Z_{L} = \frac{1}{D} \sum_{l} (2l+1) Z_{l} \,.
\label{eqnzlsum}
\end{align}
Here $Z_{l} \, (\, = \sum_{m} Z_{L}/(2l+1))$ 
is the quasiparticle weight for $l$ ($=s,p,d$) electrons,
and $D$ is the number of orbitals per atom 
($D=9$ in the present case). 
The relation allows us to interpret $Z_{l}$ as a partial quasiparticle 
weight for the electrons with orbital $l$.

\section{Numerical Results}

\subsection{Systematic change of momentum distribution functions}
%
%
%
%
\begin{figure}[b]
\begin{center}
\includegraphics[width=12cm]{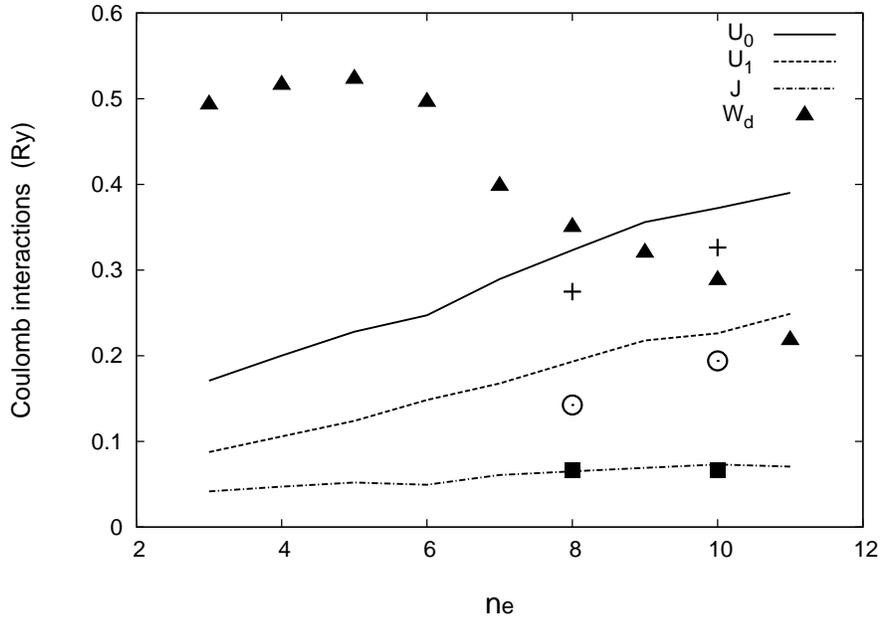}
\end{center}
\caption{
Intra-atomic Coulomb and exchange energy parameters as a function of 
the conduction electron number $n_{e}$ of iron-group transition metals.
These parameters are obtained from the
band~\cite{bdyo89} and atomic~\cite{mann67} calculations.  
Intra-orbital Coulomb interactions $U_{0}$: solid curve,
inter-orbital Coulomb interactions $U_{1}$: dashed curve,
exchange interactions $J$: dot-dashed curve.
$U_{0}$, $U_{1}$, and $J$ used by Anisimov {\it et al.}~\cite{anis97-2} 
are also shown by $+$, $\odot$, and $\blacksquare$ 
for Fe ($n_{e}=8$) and Ni ($n_{e}=10$). 
Closed triangles $\blacktriangle$ indicate the $d$ band width 
$W_{d}$~\cite{ander85}. 
}
\label{figuj}
\end{figure}
%
%

In the calculations of the momentum distribution function
(MDF) for the iron-group transition metals, 
we adopted the same lattice constants and
structures as used by Andersen {\it et al.}~\cite{ander85}, 
and constructed the tight-binding LDA+U Hamiltonians using
the Barth-Hedin exchange-correlation potential~\cite{barth72}.  
Furthermore we assumed orbital-independent Coulomb and exchange 
interactions 
$U_{mm}=U_{0}$, $U_{mm'}=U_{1}$ ($m' \neq m$), and $J_{mm'}=J$. 
These values are obtained from the average Coulomb interaction 
energies $U$ via the relations $U_{0}=U+8J/5$ and $U_{1}=U-2J/5$ 
for the cubic system.
We applied the average interactions $U$ obtained by Bandyopadhyay 
{\it et al.}~\cite{bdyo89} 
and the average $J$ obtained from the Hartree-Fock 
atomic calculations~\cite{mann67}. 
The Coulomb and exchange interaction energies from Sc and Cu are 
depicted in Fig. \ref{figuj} as a function of the conduction 
electron number $n_{e}$.
The same Hamiltonian and Coulomb-exchange interactions have been 
applied in the investigations of the excitation spectra in 3$d$
transition metals with use of the first-principles 
DCPA~\cite{kake10}.

We performed the self-consistent Hartree-Fock calculations from Sc
to Cu in the paramagnetic state using the tight-binding 
LDA+U Hamiltonian.  With use of the Hartree-Fock energy 
bands and eigenvectors, 
we solved the self-consistent equations (\ref{eqcorr}), 
(\ref{eqsclam}), and (\ref{eqne}), 
%
%
\begin{figure}
\begin{center}
\includegraphics[width=12cm]{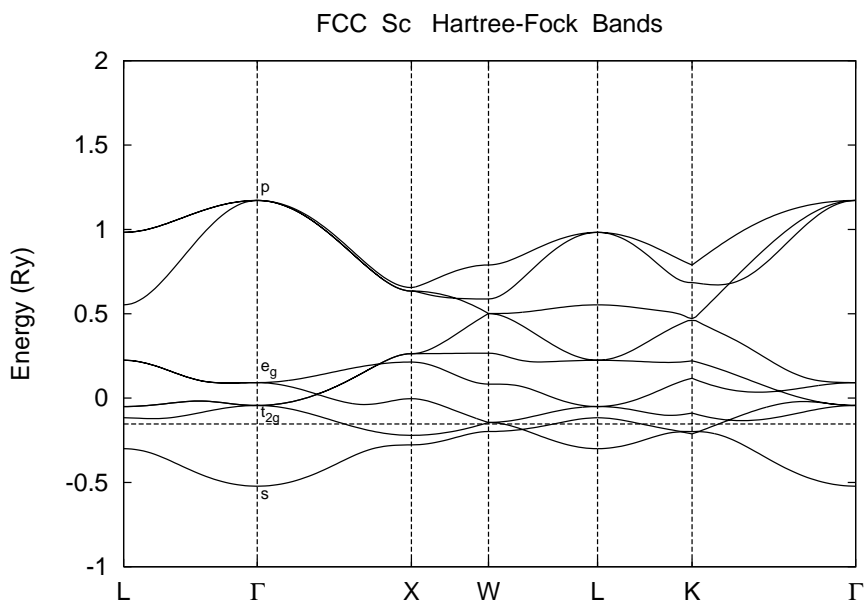}
\end{center}
\caption{Hartree-Fock one-electron energy bands of fcc Sc along 
high-symmetry lines of the first Brillouin zone.  The Fermi level 
($-0.1538$ Ry) is expressed by the horizontal dashed line.
Point symmetries of eigen functions at the $\Gamma$ point are expressed 
by $s, \, p, \, e_{g}$, and $t_{2g}$. 
}
\label{figekhfsc}
\end{figure}
%
%
and calculated the momentum distribution
functions (MDF) from Sc to Cu according to Eq. (\ref{eqnkn}).  

As has been discussed in the last paper~\cite{kake16-2}, the $d$ band
widths in the Hartree-Fock approximation are broader than the LDA ones
for the elements with $d$ electrons less than half by about 10-30 \%,
while they shrink for the elements with $d$ electrons more than half by
several percent.
We show the calculated Hartree-Fock energy bands of fcc Sc along 
high-symmetry lines 
in Fig. \ref{figekhfsc}.
%
%
\begin{figure}[htb]
\begin{center}
\includegraphics[width=12cm]{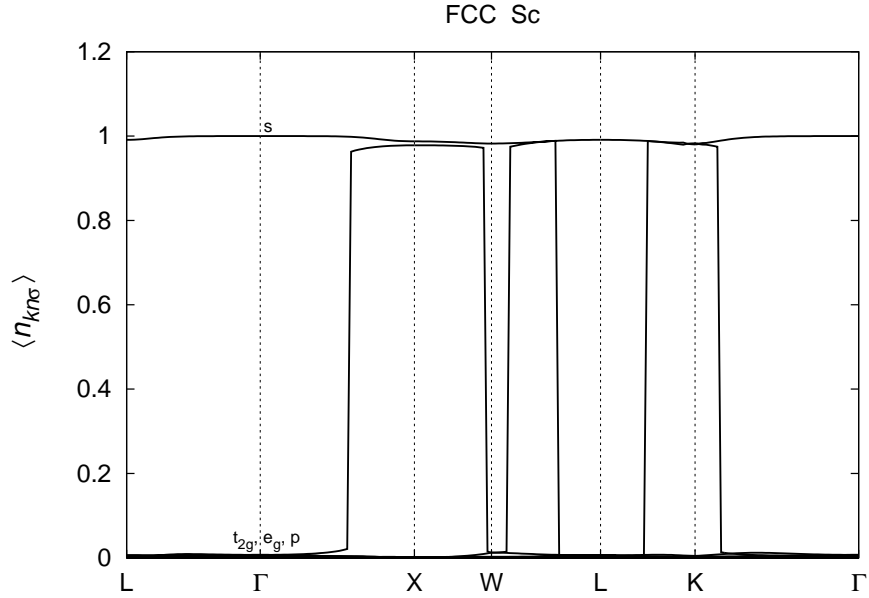}
\end{center}
\caption{Momentum distribution functions 
$\langle n_{kn\sigma} \rangle$ along high-symmetry lines for fcc Sc.  
The branches at the $\Gamma$ point are shown by their orbital 
symmetries (,{\it i.e.}, $s, \, p, \, e_{g}$, and $t_{2g}$).
}
\label{fignknsc}
\end{figure}
%
%
%
%
\begin{figure}[h]
\begin{center}
\includegraphics[width=10cm]{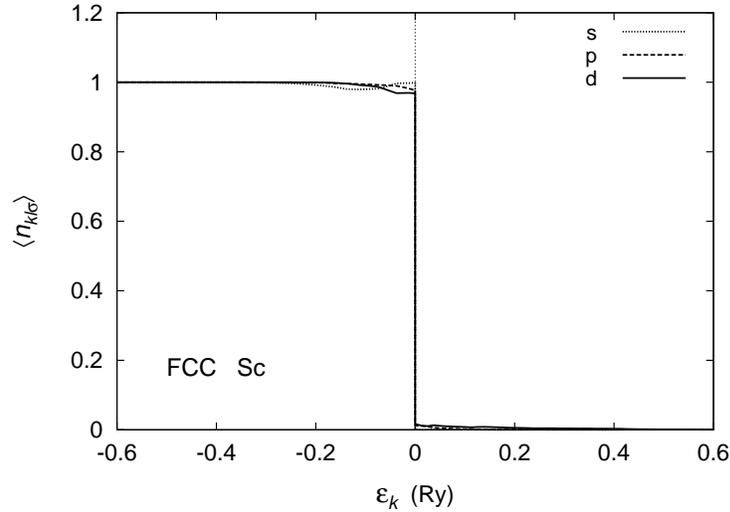}
\end{center}
\caption{The projected momentum distribution functions (MDF)
$\langle n_{kl\sigma}\rangle$ as a function of the energy
 $\epsilon_{k}$ ( = $\epsilon_{kL\sigma} - \epsilon_{\rm F}$) 
for fcc Sc. 
Dotted curve: the MDF for $s$ electrons ($l=0$), dashed curve: the MDF 
for $p$ electrons ($l=1$), solid curve: the MDF for $d$ electrons 
($l=2$).}
\label{fignklsc}
\end{figure}
%
%
There are 4 eigenvalues at point $\Gamma$: $-0.522$ Ry for $s$ electrons
below the Fermi level $\epsilon_{\rm F} \,(= -0.154 \,{\rm Ry})$,
$-0.044$ Ry for $t_{2g}$ electrons above $\epsilon_{\rm F}$, 
0.092 Ry for $e_{g}$ electrons, and
1.172 Ry for $p$ electrons.  When the wavevector $\boldsymbol{k}$ 
moves to point
$X$ along the $\Gamma$-X line, the energy for $s$ electrons below
$\epsilon_{\rm F}$ increases, hybridizes with $e_{g}$ 
electrons, and
has a value $-0.277$ Ry at point X.  The energy band for $t_{2g}$
electrons above $\epsilon_{\rm F}$ splits into two branches with the 
change of $\boldsymbol{k}$ towards point X.  
One decreases, crosses the Fermi level at 
$\boldsymbol{k}_{\rm F} = (0, 0.58, 0)$ 
in the unit of $2\pi/a$, $a$ being 
the lattice constant, and takes a value $-0.220$ Ry with 
the $xz$ symmetry at point X.  
%
%
\begin{figure}[h]
\begin{center}
\includegraphics[width=12cm]{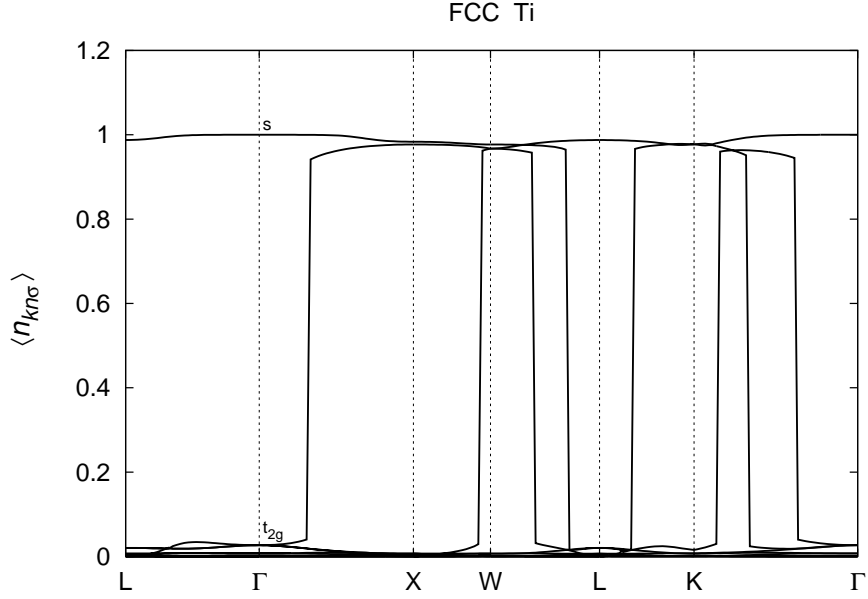}
\end{center}
\caption{Momentum distribution functions 
$\langle n_{kn\sigma} \rangle$ along high-symmetry lines for fcc Ti.  
}
\label{fignknti}
\end{figure}
%
%
%
%
\begin{figure}[h]
\begin{center}
\includegraphics[width=10cm]{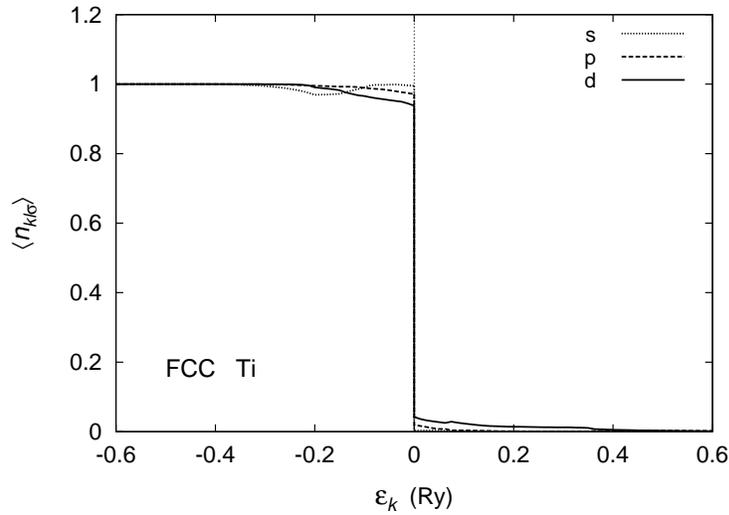}
\end{center}
\caption{The projected MDF $\langle n_{kl\sigma}\rangle$ as a function 
of the energy $\epsilon_{k}$ for fcc Ti. 
Dotted curve: the MDF for $s$ electrons, 
dashed curve: the MDF for $p$ electrons, solid curve: the MDF 
for $d$ electrons.}
\label{fignklti}
\end{figure}
%
%
Another is two-fold degenerate, and increases, takes a value 0.263
Ry at point X.  
The energy band for $e_{g}$ electrons splits into two branches on the
$\Gamma$-X line.  One decreases with the change of $\boldsymbol{k}$ 
towards point X, and takes a value $-0.004$ Ry with $y$ symmetry 
at point X.
Another gradually increases with the change of $\boldsymbol{k}$, 
and takes a value 0.213 Ry with $e_{g}$ symmetry at point X.
Note that the MDF in the Hartree-Fock approximation takes the value 1
for occupied electrons below $\epsilon_{\rm F}$ and the value 0 for
unoccupied electrons above $\epsilon_{\rm F}$, and jumps at the Fermi
surface by $\pm 1$.

We present in Fig. \ref{fignknsc} the MDF for fcc Sc.
The MDF for $s$ electrons with energy below $\epsilon_{\rm F}$ has 
a value 1.000 
at point $\Gamma$ in agreement with the result of the Fermi distribution
function (FDF), {\it i.e.}, $f(\tilde{\epsilon}_{kn\sigma})$.  
It slightly decreases when the wavevector $\boldsymbol{k}$
moves to point X along the $\Gamma$-X line, and takes a value 0.988 
at point X, which is slightly smaller than the Hartree-Fock value 1, 
because of the electron correlations of $e_{g}$ electrons via 
hybridization between the $s$ and $e_{g}$ electrons.  
The MDF for $t_{2g}$ electrons 
with $xz$ symmetry above $\epsilon_{\rm F}$ shows a small value 0.007 at
point $\Gamma$ and increases with the change of $\boldsymbol{k}$ towards
point X, jumps up from 0.021 to 0.963 at 
$\boldsymbol{k}_{\rm F} = (0, 0.58, 0)$, 
and increases further.  
The quasiparticle weight at $\boldsymbol{k}_{\rm F} = (0, 0.58, 0)$ 
has a value $Z_{kn} = 0.942$, thus the mass enhancement factor 
$m^{\ast}_{kn}/m = 1.062$. The MDF has a value 0.978 at point X.  
The MDF for the other $t_{2g}$ electrons
decreases from 0.007 to 0.000 with the change of $\boldsymbol{k}$ on the
$\Gamma$-X line.
The MDF for $e_{g}$ and $p$ electrons with energy 
above $\epsilon_{\rm F}$ have the
value 0.000 in agreement with the result of the Hartree-Fock FDF.  
Similar behavior is found also in the MDF along the X-W-L-K-$\Gamma$ 
lines.
%
%
\begin{figure}[h]
\begin{center}
\includegraphics[width=12cm]{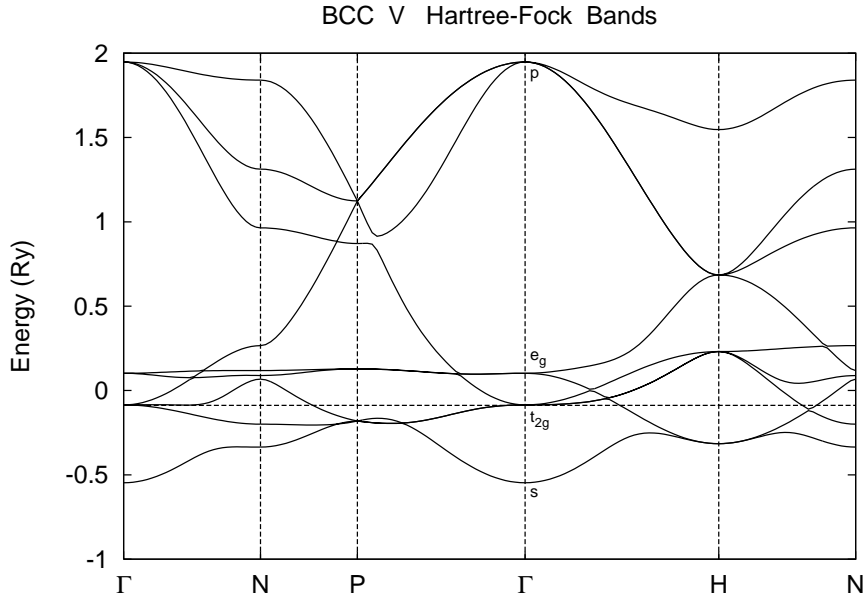}
\end{center}
\caption{Hartree-Fock one-electron energy bands of bcc V along 
high-symmetry lines.  The Fermi level 
($-0.0884$ Ry) is expressed by the horizontal dashed line.
}
\label{figekhfv}
\end{figure}
%
%

In order to make clearer the role of the $s$, $p$, and $d$ electrons, we
calculated the projected MDF for Sc as shown in Fig. \ref{fignklsc}.  
We verify that the MDF for $s$, $p$, and $d$ electrons approximately 
follow the FDF in the case of Sc, 
and thus behave as the independent electrons, 
though the MDF for $d$ electrons show a
small deviation from the FDF near the Fermi
levels ($\epsilon_{k} = 0$).

The band structure of fcc Ti is basically the same as the fcc Sc.
The Fermi level relatively shifts up and the $d$ bands are filled
more because of larger conduction electron number $n_{e}$.
Accordingly, the deviation of the MDF from the FDF 
becomes larger in fcc Ti as shown in Fig. \ref{fignknti}.  
The MDF for $t_{2g}$ electrons has
a larger value 0.027 at point $\Gamma$, and jumps up at 
$\boldsymbol{k}_{\rm F} = (0, 0.33, 0)$ 
when the wavevector $\boldsymbol{k}$
moves to point X.  The quasiparticle weight obtained
there has smaller value $Z_{kn} = 0.901$ as compared with the value
$Z_{kn} = 0.942$ in Sc, thus larger mass enhancement 
$m_{kn}/m = 1.109$.
Another branch of $t_{2g}$ electrons decreases with the change of
$\boldsymbol{k}$ towards point X, and has a value 0.007 at point X.
The MDF curve for $t_{2g}$ electrons on the $\Gamma$-L line shows a
broad peak around $\boldsymbol{k} = (0.26, 0.26, 0.26)$ because the
correponding $\epsilon_{kn}$ shows the minimum there.
The $s$ electrons with energy below $\epsilon_{\rm F}$ and 
$p$ electrons above 
$\epsilon_{\rm F}$ behave as the independent electrons as in the case of
Sc.  

Figure \ref{fignklti} shows the projected MDF for fcc Ti.  The MDF for
$s$ electrons show a small dip around $\epsilon_{k} = -0.2$ Ry due to
electron correlations via the $sd$ hybridization.  
The MDF for $p$ electrons also shows a small momentum
dependence for the same reason.
The MDF for $d$ electrons shows larger momentum dependence.  Calculated
partial mass enhancement factors are $m^{\ast}_{s}/m = 1.008$, 
$m^{\ast}_{p}/m = 1.051$, and $m^{\ast}_{d}/m = 1.117$, respectively.
%
%
\begin{figure}[h]
\begin{center}
\includegraphics[width=12cm]{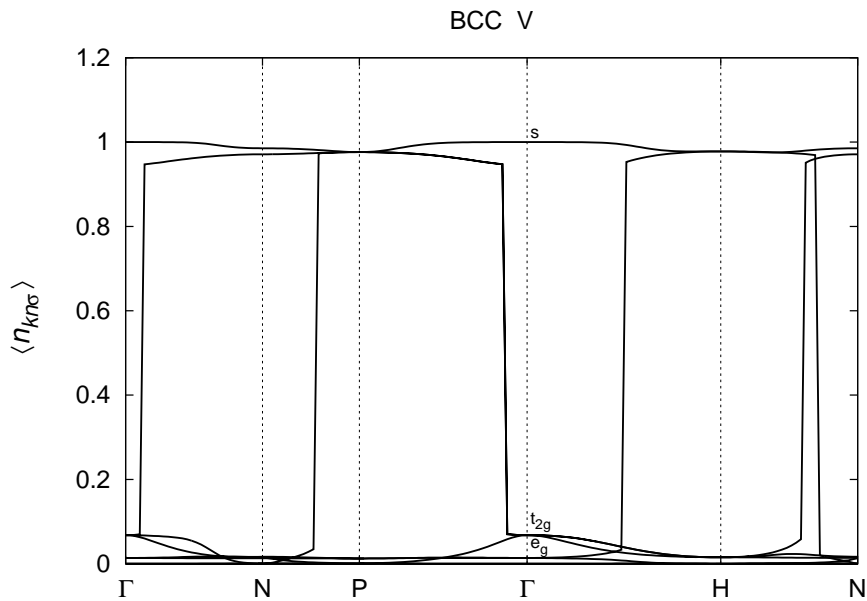}
\end{center}
\caption{Momentum distribution functions 
$\langle n_{kn\sigma} \rangle$ along high-symmetry lines for bcc V.  
}
\label{fignknv}
\end{figure}
%
%
%
%
\begin{figure}[h]
\begin{center}
\includegraphics[width=10cm]{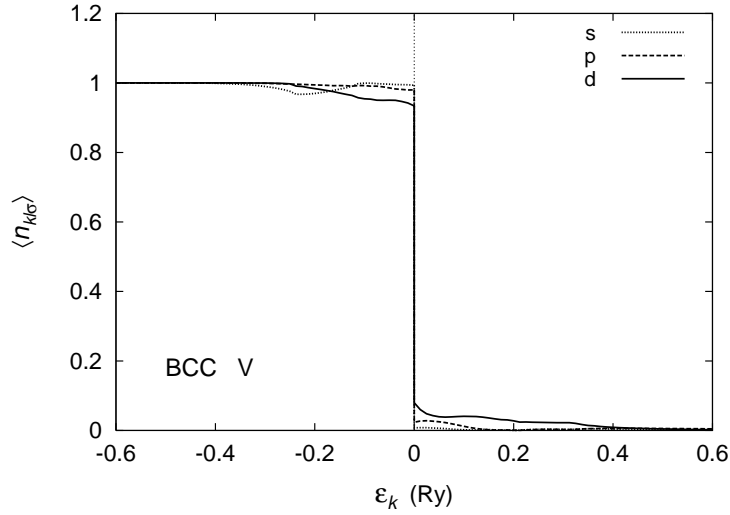}
\end{center}
\caption{The projected MDF $\langle n_{kl\sigma}\rangle$ as a 
function of the energy $\epsilon_{k}$ for bcc V. 
Dotted curve: the MDF for $s$ electrons, dashed curve: the MDF for $p$
 electrons, solid curve: the MDF for $d$ electrons.}
\label{fignklv}
\end{figure}
%
%

The Hartree-Fock band structure for bcc V is presented in Fig. 
\ref{figekhfv}.  The energy eigenvalues at point $\Gamma$ are 
$-0.547$ Ry for $s$ electrons below $\epsilon_{\rm F}$, $-0.085$ Ry 
for $t_{2g}$ electrons just above $\epsilon_{\rm F} (=-0.088 \ {\rm Ry})$, 
0.102 Ry for $e_{g}$ electrons
above $\epsilon_{\rm F}$, and 1.948 Ry for $p$ electrons far above 
$\epsilon_{\rm F}$.
When the wavevector $\boldsymbol{k}$ moves to point H along the
$\Gamma$-H line, the energy bands for $e_{g}$ and $t_{2g}$ electrons
split into two branches respectively.  Among them, one of the branches
for $e_{g}$ electrons crosses the Fermi level at 
$\boldsymbol{k}_{\rm F} = (0, 0.49, 0)$, 
where the Hartree-Fock MDF changes the value from 0 to 1 
according to the FDF.
  
The MDF curves of bcc V for correlated electrons are presented in
Fig. \ref{fignknv}.  There are 4 different MDF at point $\Gamma$ 
whose values are 1.000 for $s$ electrons with energy 
below $\epsilon_{F}$, 0.068 for
$t_{2g}$ electrons just above $\epsilon_{\rm F}$, 0.014 for $e_{g}$
electrons above $\epsilon_{\rm F}$, and 0.000 for $p$ electrons,
respectively. 
When the wavevector $\boldsymbol{k}$ moves to point H along the
$\Gamma$-H line, the MDF for $s$ electrons slightly decreases due to
the $sd$ hybridization and has a value 0.977 at point H.
The MDF for $t_{2g}$ electrons splits into two branches.  Both branches
gradually decrease with the change of $\boldsymbol{k}$ towards point H 
and again merge
into a single band with the value 0.015 at point H.  The MDF for
$e_{g}$ electrons also splits into two branches.  One monotonically
decreases and reduces to zero at point H.  Another branch monotonically
increases with the change of $\boldsymbol{k}$, jumps up at 
$\boldsymbol{k}_{\rm F} = (0, 0.50, 0)$ from 0.033 to 0.953, 
and has a value
0.977 at point H.  The MDF for $p$ electrons are almost zero on the
$\Gamma$-H line because the corresponding energies $\epsilon_{kn}$ are
far above $\epsilon_{\rm F}$.  
The jumps at $\boldsymbol{k}_{\rm F} = (0.50, 0.50, 0.09)$ along
the $\Gamma$-N line and 
$\boldsymbol{k}_{\rm F} = (0.07, 0.07, 0.07)$ along the 
P-$\Gamma$ line yield the smallest quasiparticle weight $Z_{kn}$, thus
the largest mass enhancement factor $m^{\ast}_{kn}/m = 1.140$ which is
larger than that of the fcc Ti.

In Fig. \ref{fignklv}, we show the projected MDF for bcc V.  We find
that the basic behavior is similar to that in the fcc Ti (see
Fig. \ref{fignklti}). However the deviation of the MDF for $d$ 
electrons from the FDF becomes larger.

The Hartree-Fock band structure of bcc Cr is similar to the bcc V.  The
Fermi level is shifted up by about 0.06 Ry, so that new Fermi
surfaces appear at $\boldsymbol{k}_{\rm F} = (0.20, 0.20, 0)$ on the
$\Gamma$-N line, $\boldsymbol{k}_{\rm F} = (0, 0.37, 0)$ and
$(0, 0.60, 0)$ on the $\Gamma$-H line.
%
%
\begin{figure}[h]
\begin{center}
\includegraphics[width=12cm]{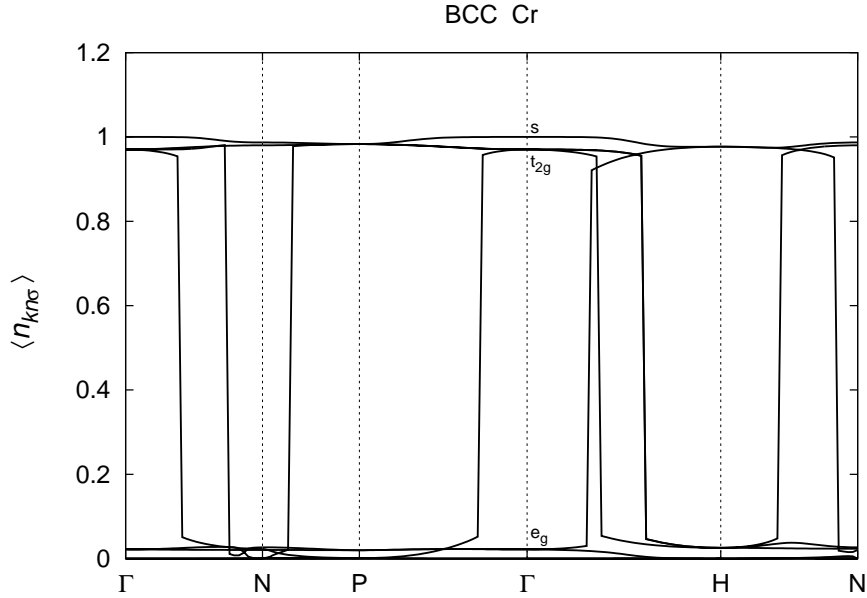}
\end{center}
\caption{Momentum distribution functions 
$\langle n_{kn\sigma} \rangle$ along high-symmetry lines for bcc Cr. 
}
\label{fignkncr}
\end{figure}
%
%
%
%
\begin{figure}[h]
\begin{center}
\includegraphics[width=10cm]{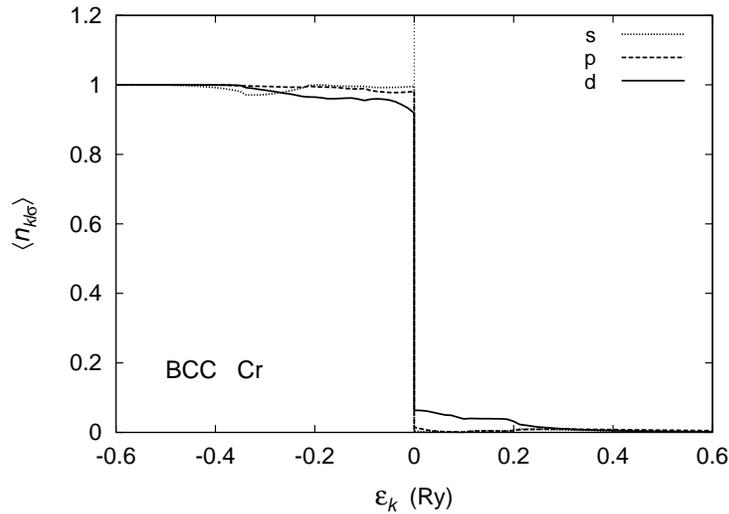}
\end{center}
\caption{The projected MDF $\langle n_{kl\sigma}\rangle$ as a function 
of the energy $\epsilon_{k}$ for bcc Cr. 
Dotted curve: the MDF for $s$ electrons, 
dashed curve: the MDF for $p$ electrons, solid curve: 
the MDF for $d$ electrons.}
\label{fignklcr}
\end{figure}
%
%
Accordingly new jumps of the MDF appear at these $\boldsymbol{k}$
points as shown in Fig. \ref{fignkncr}.  
The deviations of the MDF from the FDF 
are comparable to those in the bcc V.  Calculated projected MDF also
show the behavior similar to the bcc V as shown in Fig. \ref{fignklcr},
though the dip of the MDF for $s$ electrons due to the $sd$
hybridization is now located around $\epsilon_{k} = -0.3$ Ry.  

\begin{figure}[h]
\begin{center}
\includegraphics[width=12cm]{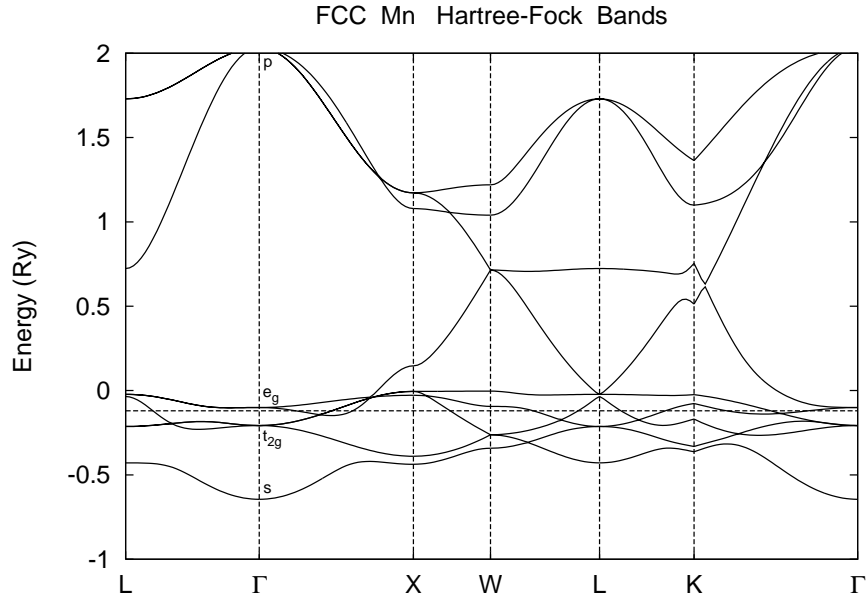}
\end{center}
\caption{Hartree-Fock one-electron energy bands of fcc Mn along 
high-symmetry lines.  The Fermi level 
($-0.1201$ Ry) is expressed by the horizontal dashed line.
}
\label{figekhfmn}
\end{figure}

Next, we discuss the MDF for fcc Mn.  The fcc Mn has larger Coulomb and
exchange interactions as shown in Fig. \ref{figuj}.
The band structure is shown in
Fig. \ref{figekhfmn}.  The $e_{g}$ bands sink more with increasing 
the electron
number $n_{e}$, and are located on the Fermi level.
Because the $e_{g}$ bands are narrow and the
$t_{2g}$ bands are also located near the Fermi level, calculated MDF for
$d$ electrons are expected to show a large deviation from the FDF.
%
%
%
%

\begin{figure}[h]
\begin{center}
\includegraphics[width=12cm]{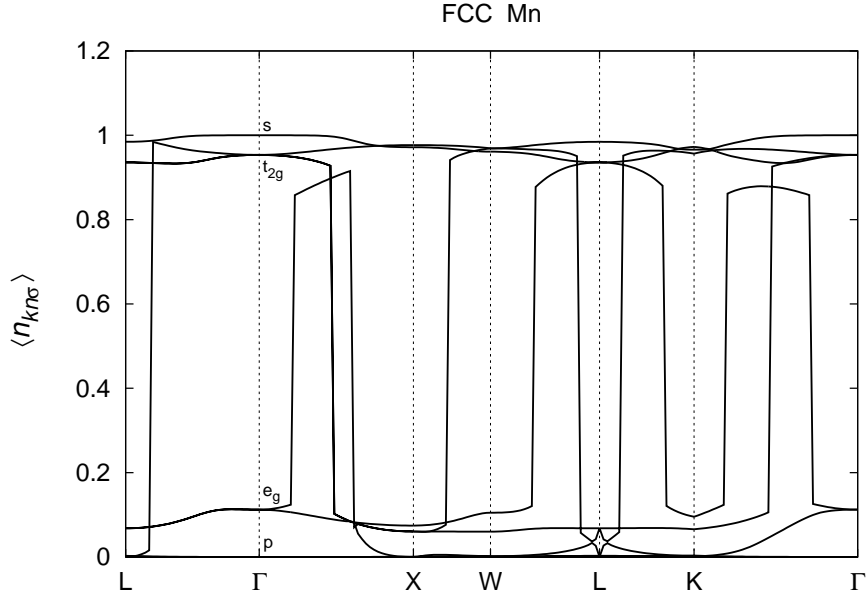}
\end{center}
\caption{Momentum distribution functions 
$\langle n_{kn\sigma} \rangle$ along high-symmetry lines for fcc Mn. 
}
\label{fignknmn}
\end{figure}

Figure \ref{fignknmn} shows the MDF for fcc Mn along high-symmetry
lines.  The MDF for $s$
band has a value 1.000 at point $\Gamma$.  When the wavevector
$\boldsymbol{k}$ moves to point X along the $\Gamma$-X line, it
monotonically decreases due to hybridization between $s$ and $t_{2g}$
electrons, and has a value 0.970 at point X.  
The MDF for $d$ electrons show distinct deviation from the FDF. 
The MDF for $t_{2g}$
electrons with energy below $\epsilon_{\rm F}$ has a value 0.954 
at point $\Gamma$.  
With the change of $\boldsymbol{k}$ towards point X 
it splits into two branches.  The upper branch with
$xy$ symmetry slightly increases and has a value 0.977 at point X.  The
lower branch decreases along the $\Gamma$-X line, jumps down at
$\boldsymbol{k}_{\rm F} = (0, 0.47, 0)$, and 
reaches the X point.  It has a value 0.060 at point X.
The MDF for $e_{g}$ electrons has a value 0.112 at point $\Gamma$.  It
splits into two branches along the $\Gamma$-X line.  The first branch
monotonically decreases with the change of $\boldsymbol{k}$ and takes a
value 0.074 at point X.  The second one increases first, jumps up at
$\boldsymbol{k}_{\rm F} = (0, 0.22, 0)$, and increases further along the
$\Gamma$-X line.  But it again jumps down at 
$\boldsymbol{k}_{\rm F} = (0, 0.59, 0)$.  
Finally it decreases and has the
value 0.000 with the $p$ symmetry at point X.     
%
%
%
%
%
%
\begin{figure}[tbh]
\begin{center}
\includegraphics[width=10cm]{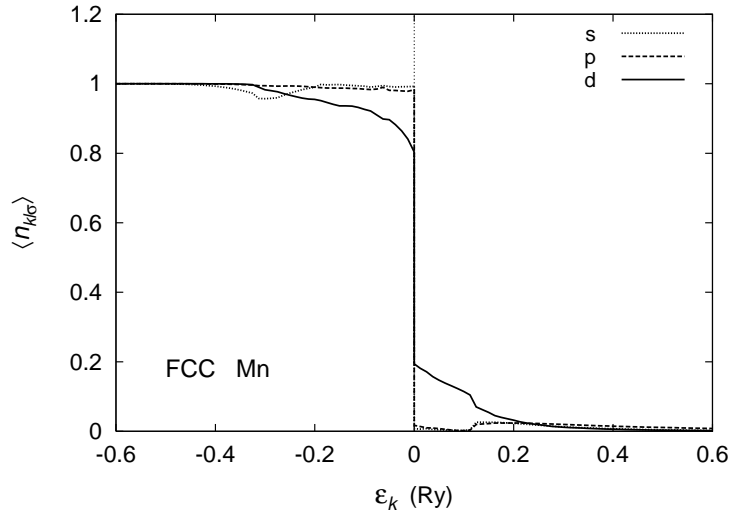}
\end{center}
\caption{The projected MDF $\langle n_{kl\sigma}\rangle$ as a function 
of the energy $\epsilon_{k}$ for fcc Mn. 
Dotted curve: the MDF for $s$ electrons, 
dashed curve: the MDF for $p$ electrons, solid curve: 
the MDF for $d$ electrons.}
\label{fignklmn}
\end{figure}
%
%
The MDF for $e_{g}$ electrons show more significant deviations from the
FDF than those for $t_{2g}$ case, 
and show a considerable mass enhancement on the Fermi
surface: $m^{\ast}_{kn}/m = 1.362$ at 
$\boldsymbol{k}_{\rm F} = (0, 0.22, 0)$ on the $\Gamma$-X line and 
$m^{\ast}_{kn}/m = 1.365$ at 
$\boldsymbol{k}_{\rm F} = (0, 0.21, 0.21)$ on the K-$\Gamma$ line.

The projected MDF for fcc Mn show a clear difference between the $d$
electrons and $sp$ electrons as seen in Fig. \ref{fignklmn}.
The MDF for $d$ electrons shows a strong momentum dependence due to
electron correlations, while those for $sp$ electrons are rather close
to the FDF.
We find the partial mass enhancement factors, $m^{\ast}_{s}/m = 1.015$,
$m^{\ast}_{p}/m = 1.035$, and $m^{\ast}_{d}/m = 1.640$, respectively.   
%
%
\begin{figure}[tbh]
\begin{center}
\includegraphics[width=12cm]{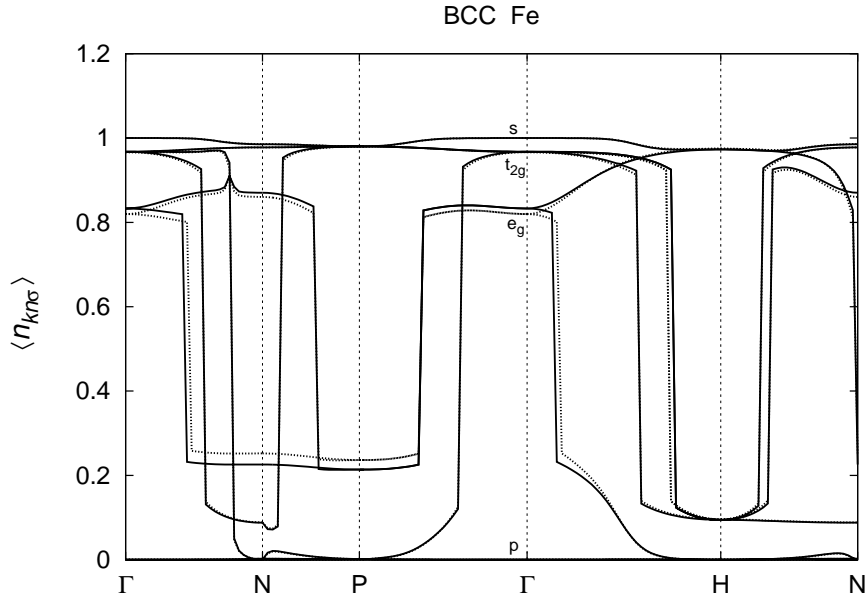}
\end{center}
\caption{Momentum distribution functions 
$\langle n_{kn\sigma} \rangle$ along high symmetry lines for bcc Fe.  
Dotted curves are the result~\cite{kake16} with use of $U=0.1691$ Ry and 
$J=0.0662$ Ry~\cite{anis97-2}.
}
\label{fignknfe}
\end{figure}
%
%

We have discussed the MDF for bcc Fe with use of the average 
Coulomb and exchange 
interactions $U = 0.1691$ Ry and $J = 0.0662$ Ry~\cite{anis97-2} 
recently~\cite{kake16}.
We present here the results for $U = 0.2192$ Ry and $J = 0.0650$ Ry 
obtained by Bandyopadhyay {\it et al.} and the Hartree-Fock
atomic calculations as mentioned before.
Figure \ref{fignknfe} shows the MDF for Fe along high-symmetry
lines.  As seen from the figure, the difference in the MDF 
between two sets 
of $U$ and $J$ is small.  
At point $\Gamma$, we have 4 branches of the MDF.  The MDF
for $s$ electrons with energy eigen value $-0.687$ Ry 
below $\epsilon_{\rm F}$ ($= -0.134$ Ry) has a value
1.000 at point $\Gamma$.  It monotonically decreases when the 
wavevector $\boldsymbol{k}$ moves to point H along the $\Gamma$-H line,
and takes a value 0.973 at point H.  

The MDF for $t_{2g}$ electrons with
energy $-0.263$ Ry below $\epsilon_{\rm F}$ has a value 0.967 at point
$\Gamma$.  With the change of $\boldsymbol{k}$ towards 
point H, it
splits into two branches.  The first branch decreases with the change of
$\boldsymbol{k}$, jumps down at 
$\boldsymbol{k}_{\rm F} = (0, 0.76, 0)$, and
finally takes a value 0.095 at point H.  Another branch decreases more
rapidly, jumps down at $\boldsymbol{k}_{\rm F} = (0, 0.58, 0)$ 
and has the same value 0.095 at point H.  

The MDF for $e_{g}$ electrons with energy just below 
$\epsilon_{\rm F}$
shows the largest deviation from the FDF (=1), {\it i.e.}, 
$\langle n_{kn\sigma} \rangle = 0.832$ at point $\Gamma$ 
because the flat bands of $e_{g}$ electrons along
$\Gamma$-N-P-$\Gamma$ line are located on the Fermi level in the case of
bcc Fe (see Fig. \ref{figekhfv}).  
It splits into two branches with the change of
$\boldsymbol{k}$ towards point H.  
The first branch monotonically increases
and takes a value 0.973 at point H.  The second branch decreases,
and jumps down at $\boldsymbol{k}_{\rm F} = (0, 0.14, 0)$ from 0.823 
to 0.232.  It further decreases
with the change of the symmetry from the $e_{g}$ to $sp$ type, and
takes the value 0.000 at point H.
We find that the $e_{g}$ electrons cause a large deviation of the MDF
from the FDF because of the strong electron correlations 
in the narrow $e_{g}$ band on the Fermi level.

%
%
\begin{table}[tbh]
\caption{Mass enhancement factors of bcc Fe for $e_{g}$ electrons 
at various wave vectors $\boldsymbol{k}$ on the Fermi surface.
\vspace{5mm} }
\label{tbccfemeffeg}
\begin{tabular}{ccccc}
\hline
$\boldsymbol{k}$  & (0.22, 0.22, 0.00) & (0.50, 0.50, 0.28) & 
(0.32, 0.32, 0.32) & (0.00, 0.14, 0.00) \\ \hline 
$m^{\ast}_{kn}/m $ & 1.70 & 1.61 & 1.66 & 1.69 \\ 
\hline
\end{tabular}
\end{table}
%
%
%
%
\begin{table}[tbh]
\caption{Mass enhancement factors of bcc Fe for $t_{2g}$ electrons 
at various wave vectors $\boldsymbol{k}$ on the Fermi surface.
\vspace{5mm} }
\label{tbccfemefft2g}
\begin{tabular}{ccccc}
\hline
$\boldsymbol{k}$  & (0.28, 0.28, 0.00) & (0.39,0.39, 0.00) & 
(0.50, 0.50, 0.09) & (0.29, 0.29, 0.29)  \\ \hline 
$m^{\ast}_{kn}/m$ & 1.26 & 1.12 & 1.14 & 1.23  \\ 
\hline
\end{tabular}
\vspace*{3mm}
\begin{tabular}{ccccc}
\hline
$\boldsymbol{k}$  & (0.00, 0.58, 0.00) & (0.00, 0.76, 0.00) & 
(0.15, 0.85, 0.00) & (0.18, 0.82, 0.00) \\ \hline 
$m^{\ast}_{kn}/m$ & 1.27 & 1.25 & 1.23 & 1.26 \\ 
\hline
\end{tabular}
\end{table}
%
%
%
%
\begin{figure}[tbh]
\begin{center}
\includegraphics[width=10cm]{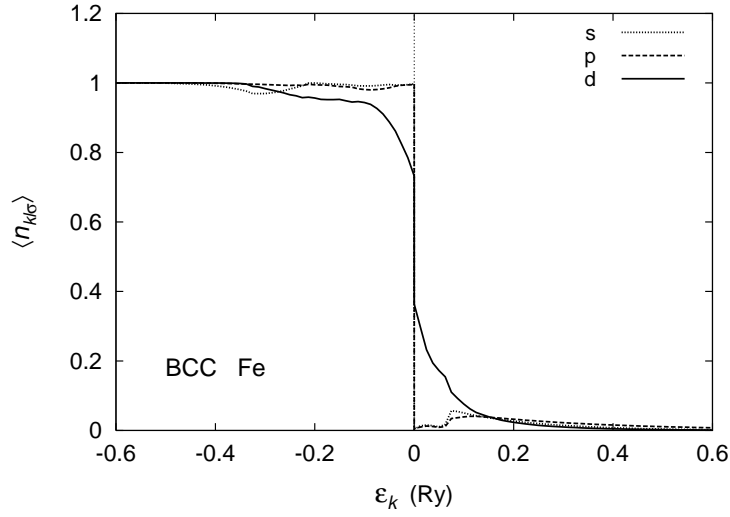}
\end{center}
\caption{The projected MDF $\langle n_{kl\sigma}\rangle$ as a 
function of the energy $\epsilon_{k}$ for bcc Fe. 
Dotted curve: the MDF for 
$s$ electrons, dashed curve: the MDF for $p$ electrons, 
solid curve: the MDF for $d$ electrons.}
\label{fignklfe}
\end{figure}
%
%
The projected MDF for $d$ electrons shows the strong momentum dependence
especially near the Fermi level as shown in Fig. \ref{fignklfe}.
From the jumps of the projected MDF at $\epsilon_{\rm F}$, we find the
partial mass enhancement factors: $m^{\ast}_{s}/m = 1.007$,
$m^{\ast}_{p}/m = 1.010$, and $m^{\ast}_{d}/m = 2.720$, respectively.
Considerable deviations of the projected MDF from the FDF are also found
for $s$ and $p$ electrons. They are caused by the hybridization
between $sp$ and $d$ electrons.

\begin{figure}[h]
\begin{center}
\includegraphics[width=12cm]{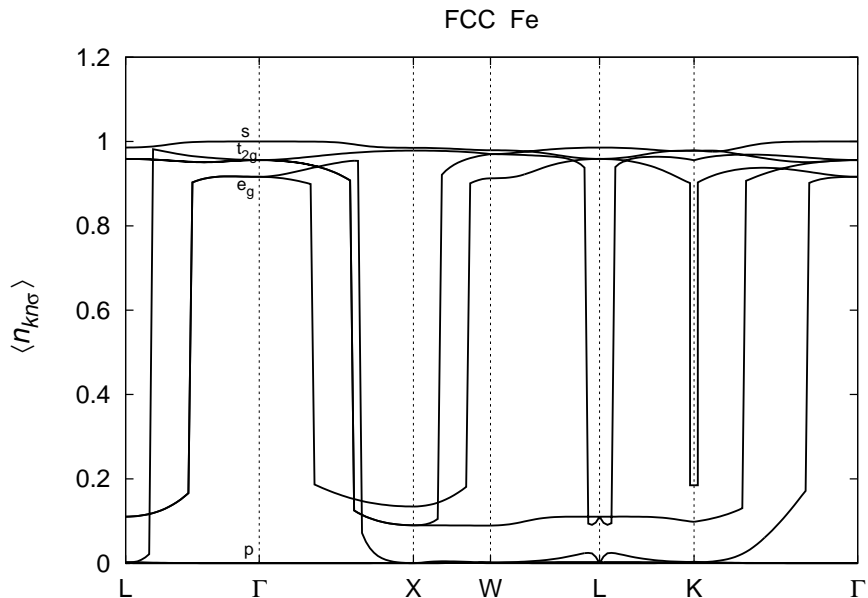}
\end{center}
\caption{Momentum distribution functions 
$\langle n_{kn\sigma} \rangle$ along high-symmetry lines for fcc Fe.  
}
\label{fignknfefcc}
\end{figure}

We also calculated the MDF for fcc Fe to clarify the
difference in the MDF between the bcc and fcc structures.
The fcc Fe shows the band structure similar to the fcc Mn (see
Fig. \ref{figekhfmn}).
Because of the change of the Fermi level, the $e_{g}$ bands of fcc Fe 
measured from $\epsilon_{\rm F}$ sink and are located just 
below $\epsilon_{\rm F}$ at
points $\Gamma$ and W, and just above $\epsilon_{\rm F}$ 
at point K.  
Accordingly, the MDF for $e_{g}$ electrons at point $\Gamma$,
W, K shift up to 0.916, 0.913, and 0.185, respectively as shown in 
Fig. \ref{fignknfefcc}, when they are
compared with those in fcc Mn (see Fig. \ref{fignknmn}).
The MDF for $t_{2g}$ electrons with energy above $\epsilon_{\rm F}$  
(,{\it e.g.}, $\langle n_{kn\sigma} \rangle = 0.956$ at point $\Gamma$) 
is somewhat shifted up in comparison with those for fcc Mn because
their band energies measured from $\epsilon_{\rm F}$ 
sink more. The other MDF bands shows the behavior similar to those 
for the fcc Mn.
%
%
%
%

We compare the mass enhancement factors of fcc Fe along high-symmetry
lines with those of the bcc Fe in Tables \ref{tbccfemeffeg}, 
\ref{tbccfemefft2g}, \ref{tfccfemeffeg}, \ref{tfccfemefft2g}. 
Although the MDF for $e_{g}$ electrons in both the fcc and bcc Fe show 
a significant deviation from the FDF, the bcc Fe shows 
larger mass enhancement on the Fermi surface. This is  
because the $e_{g}$ electrons for the bcc Fe are more localized and 
form a flat energy dispersion on the Fermi surface. 
The mass enhancement factors for $t_{2g}$ electrons approximately lead
to the same average values $m^{\ast}_{t2g}/m \approx 1.2$ for both
structures, but those of the fcc Fe show stronger momentum dependence.
%
%
\begin{figure}[t]
\begin{center}
\includegraphics[width=10cm]{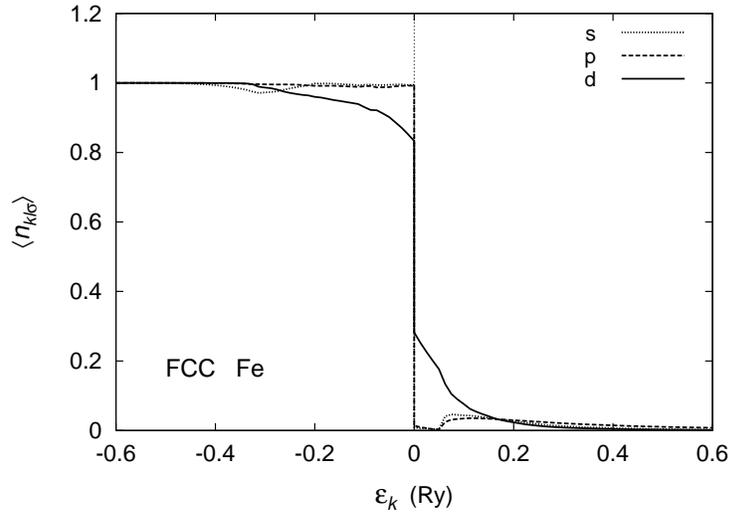}
\end{center}
\caption{The partial MDF 
$\langle n_{kl\sigma}\rangle$ as a function of the energy 
$\epsilon_{k}$ for fcc Fe. Dotted curve: the MDF for $s$ electrons, 
dashed curve: the MDF for $p$ electrons, solid curve: the MDF 
for $d$ electrons.}
\label{fignklfefcc}
\end{figure}
%
%
%
%
\begin{table}[tb]
\caption{Mass enhancement factors of fcc Fe for $e_{g}$ electrons 
at various wave vectors $\boldsymbol{k}$ on the Fermi surface.
\vspace{5mm} }
\label{tfccfemeffeg}
\begin{tabular}{ccccc}
\hline
$\boldsymbol{k}$  & (0.24, 0.24, 0.24) & (0.00, 0.35, 0.00) & 
(0.00, 0.65, 0.00) & (0.36,1.00,0.00)  \\ \hline
$m^{\ast}_{kn}/m$ & 1.36 & 1.40 & 1.13 & 1.39  \\ 
\hline
\end{tabular}
\vspace*{3mm}
\begin{tabular}{cccc}
\hline
$\boldsymbol{k}$ & (0.01,0.74,0.74) & (0.00,0.74,0.74) & 
(0.00,0.23,0.23) \\ \hline 
$m^{\ast}_{kn}/m$ & 1.40 & 1.39 & 1.37 \\ 
\hline
\end{tabular}
\end{table}
%
%
%
%
\begin{table}[tb]
\caption{Mass enhancement factors of fcc Fe for $t_{2g}$ electrons 
at various wave vectors $\boldsymbol{k}$ on the Fermi surface.
\vspace{5mm} }
\label{tfccfemefft2g}
\begin{tabular}{cccc}
\hline
$\boldsymbol{k}$  & (0.40, 0.40, 0.40) & (0.00, 0.60, 0.00) & 
(0.17, 1.00, 0.00)  \\ \hline
$m^{\ast}_{kn}/m$ & 1.04 & 1.28 & 1.23 \\ 
\hline 
\end{tabular}
\vspace*{3mm}
\begin{tabular}{cccc}
\hline
$\boldsymbol{k}$  & (0.50, 0.56, 0.44) & (0.43, 0.54, 0.54) & 
(0.00, 0.52, 0.52) \\ \hline 
$m^{\ast}_{kn}/m$ & 1.19 & 1.18 & 1.29 \\ 
\hline
\end{tabular}
\end{table}
%
%

In Fig. \ref{fignklfefcc}, we present the projected MDF for fcc Fe.
Calculated mass enhancement factors 
for $s$, $p$, and $d$ electrons are  
$m^{\ast}_{s}/m = 1.014$, $m^{\ast}_{p}/m = 1.023$, and 
$m^{\ast}_{d}/m = 1.815$, respectively.
We find that the $s$, $p$, and $d$ projected MDF curves of fcc Fe
show the similar behavior as the bcc ones.  But in the low energy region 
$|\epsilon_{k}| \lesssim 0.05$ Ry, the $d$ projected MDF of the bcc
Fe shows stronger momentum dependence leading to larger effective mass
enhancement.
%
%
\begin{figure}[b]
\begin{center}
\includegraphics[width=12cm]{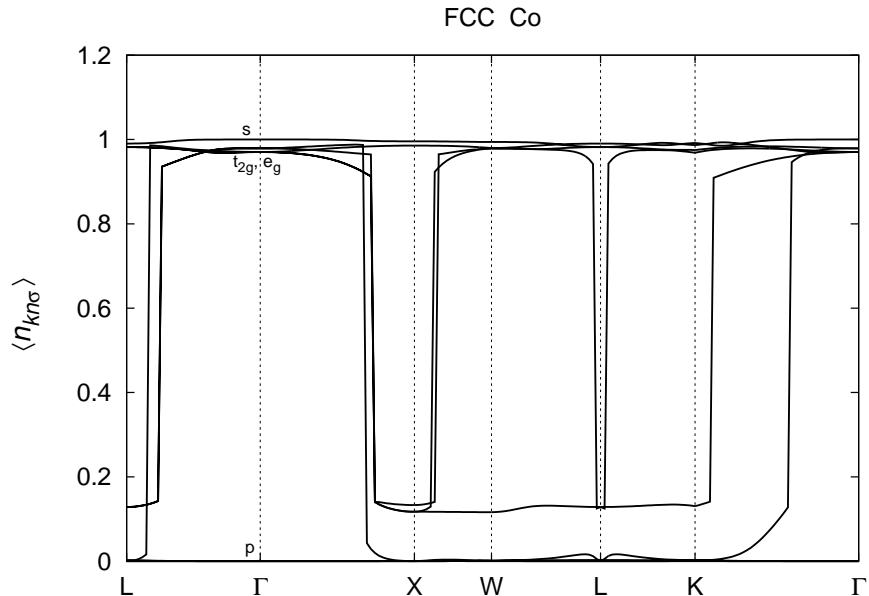}
\end{center}
\caption{Momentum distribution functions 
$\langle n_{kn\sigma} \rangle$ along high symmetry lines for fcc Co.  
}
\label{fignknco}
\end{figure}
%
%

The band structure of the fcc Co is similar to those of the fcc Mn and
fcc Fe (see Fig. \ref{figekhfmn}).  
However, the $e_{g}$ bands sink more below $\epsilon_{\rm F}$, and
the $t_{2g}$ flat bands above $\epsilon_{\rm F}$ 
along the X-W-L-K line approach to $\epsilon_{\rm F}$.  
Thus the MDF for $e_{g}$ electrons below $\epsilon_{\rm F}$ 
become closer to one as shown in Fig. \ref{fignknco}, 
and the MDF for $t_{2g}$ electrons become larger.  In fact,
the MDF for $e_{g}$ electrons at point $\Gamma$ moves from 0.916 to
0.970 when fcc Fe changes to fcc Co 
(see Figs. \ref{fignknfefcc} and \ref{fignknco}), and 
the MDF for $t_{2g}$ electrons at point $\Gamma$ moves from 0.956 
to 0.979 for the same change.  The flat MDF band for $t_{2g}$ electrons
along the X-W-L-K line has slightly enhanced value ($\approx 0.13$).
%
%
\begin{figure}[h]
\begin{center}
\includegraphics[width=10cm]{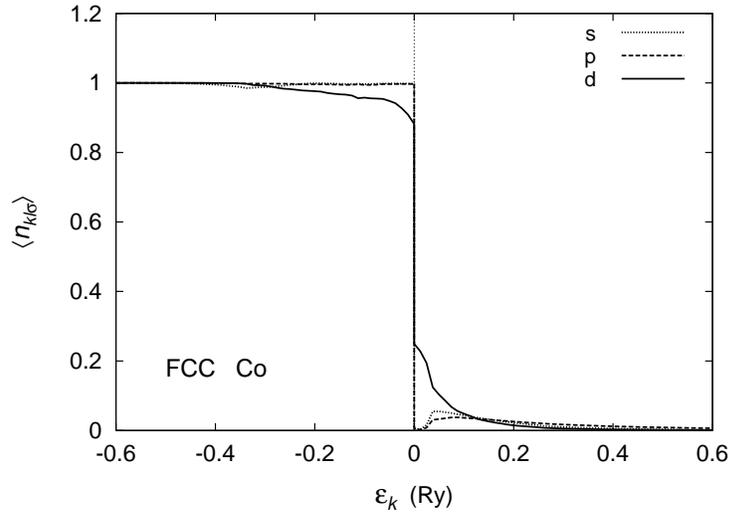}
\end{center}
\caption{The projected MDF $\langle n_{kl\sigma}\rangle$ as a function
 of the energy for fcc Co. Dotted curve: the MDF for $s$ electrons, 
dashed curve: the MDF for $p$ electrons, solid curve: the MDF 
for $d$ electrons.}
\label{fignklco}
\end{figure}
%
%
Calculated projected MDF for fcc Co are shown in Fig. \ref{fignklco}. 
The projected MDF are similar to those in the fcc Fe, but the momentum
dependence for $d$ electrons becomes weaker.
%
%
\begin{figure}[b]
\begin{center}
\includegraphics[width=12cm]{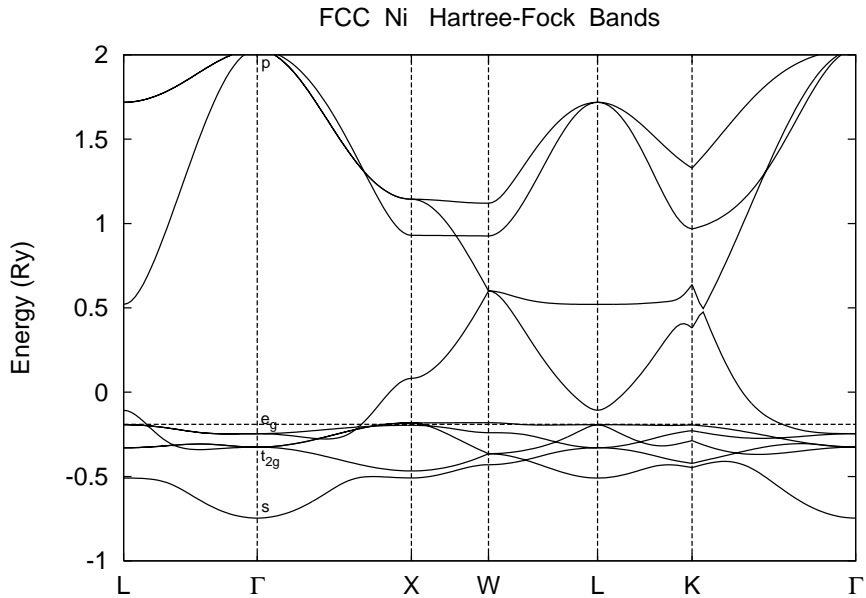}
\end{center}
\caption{Hartree-Fock one-electron energy bands of fcc Ni along 
high-symmetry lines.  The Fermi level 
($-0.1903$ Ry) is expressed by the horizontal dashed line.
}
\label{figekhfni}
\end{figure}
%
%

In the case of fcc Ni, the $d$ bands measured from the Fermi level
$\epsilon_{\rm F}$ sink further as shown in Fig. \ref{figekhfni}.
Most of the $e_{g}$ bands are located below $\epsilon_{\rm F}$.
The $t_{2g}$ branch at point $\Gamma$ is also located below
$\epsilon_{\rm F}$.  But,
the flat band of $t_{2g}$ electrons along the X-W-L-K line is on
the Fermi level.  The $sp$ bands on the other hand are located 
far below and above $\epsilon_{\rm F}$.
%
%
\begin{figure}[h]
\begin{center}
\includegraphics[width=12cm]{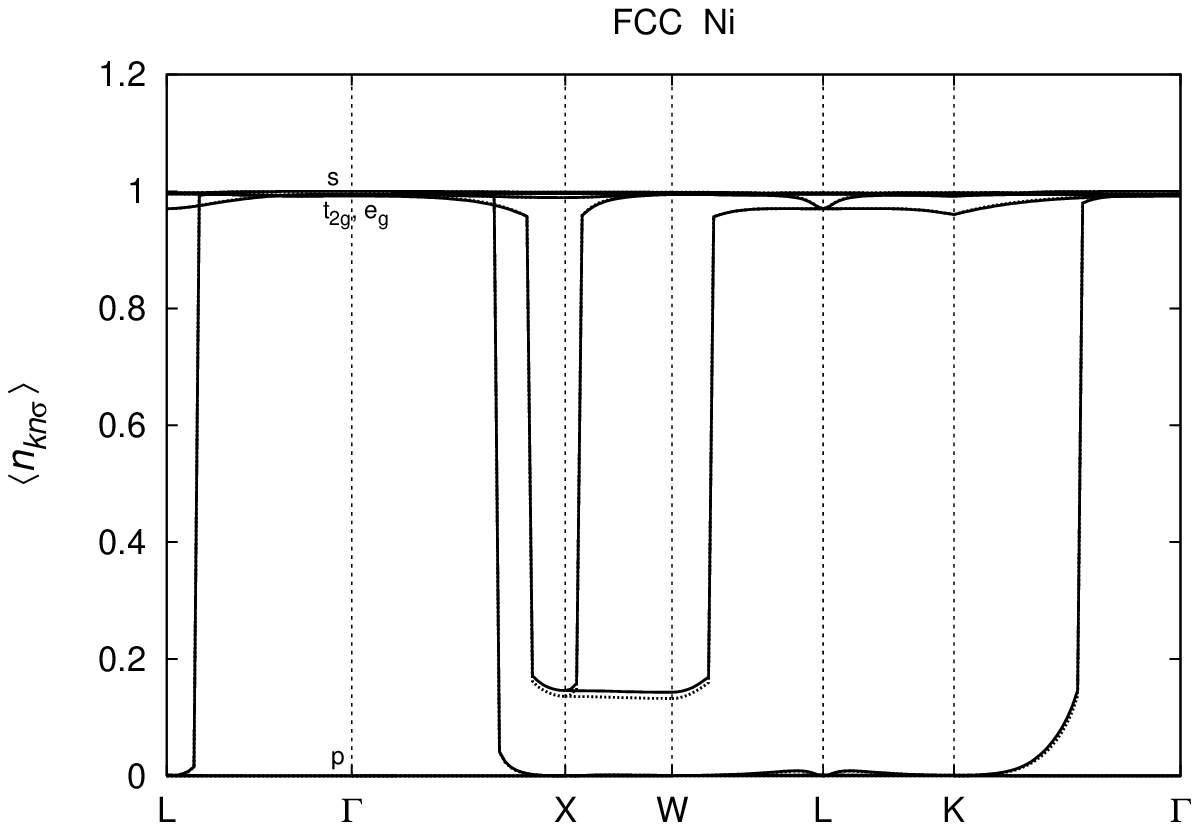}
\end{center}
\caption{Momentum distribution functions 
$\langle n_{kn\sigma} \rangle$ along high-symmetry lines for fcc Ni.  
Dotted curves are the result with use of $U=0.2205$ Ry and $J=0.0662$ Ry.
}
\label{fignknni}
\end{figure}
%
%
%
%
\begin{figure}[h]
\begin{center}
\includegraphics[width=10cm]{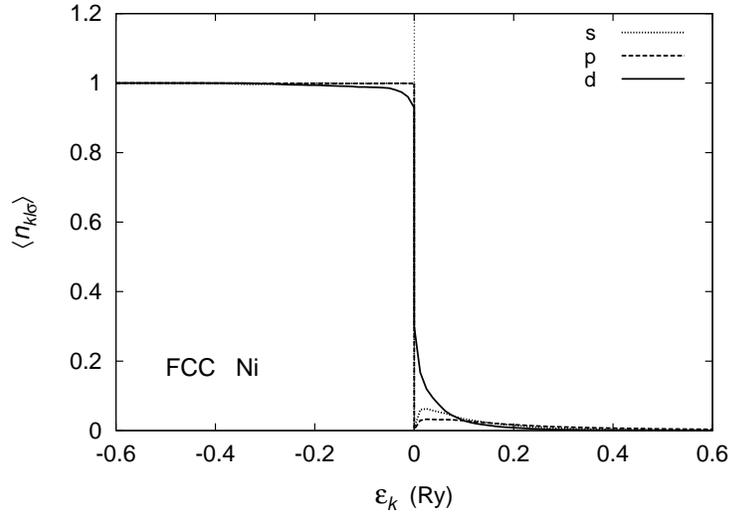}
\end{center}
\caption{The projected MDF $\langle n_{kl\sigma}\rangle$ as a function 
of the energy $\epsilon_{k}$ for fcc Ni. 
Dotted curve: the MDF for $s$ electrons, dashed curve: 
the MDF for $p$ electrons, solid curve: the MDF for $d$ electrons.}
\label{fignklni}
\end{figure}
%
%

Figure \ref{fignknni} shows the MDF for fcc Ni.  We also calculated the
MDF with use of $U = 0.2205$ Ry and $J = 0.0662$ Ry adopted by Anisimov
{\it et al.}~\cite{anis97-2} (see the dotted curves).  
We find that the difference
between the two results is small. 
Because of the band structure mentioned above, the MDF of fcc Ni at
point $\Gamma$ takes the values close to 1 or 0: 1.000 for $s$, 0.994
for $e_{g}$, 0.992 for $t_{2g}$, and 0.000 for $p$ symmetry electrons.
The MDF for $e_{g}$ electrons splits into two branches along
the $\Gamma$-X line.  The first branch hardly changes with the change 
of $\boldsymbol{k}$ and takes the
value 0.990 at point X.  The second one also hardly shows the
$\boldsymbol{k}$ dependence, but it jumps down at 
$\boldsymbol{k}_{\rm F} = (0, 0.68, 0)$ and becomes zero at point X.

The MDF for $t_{2g}$ electrons also splits into two branches when the
wavevector $\boldsymbol{k}$ moves to point X.  The branch with $xy$
symmetry remains unchanged and has a value 0.996 at point X.  The
second branch monotonically decreases with increasing 
$|\boldsymbol{k}|$ along the $\Gamma$-X line.  It jumps down at
$\boldsymbol{k}_{\rm F} = (0, 0.83, 0)$, and takes a value 0.146 at
point X. 
We find considerably large deviations from the FDF for $t_{2g}$
electrons  along the X-W-L-K line as expected from the energy band
structure. 
In particular, the $t_{2g}$ flat energy band just above
$\epsilon_{\rm F}$ on the X-W line causes a large deviation of 
the MDF.
Accordingly, the projected MDF for $d$ electrons just above
$\epsilon_{\rm F}$ shows a large deviation from the FDF as shown in
Fig. \ref{fignklni}.   A deviation from the FDF 
is also found above $\epsilon_{\rm F}$ for $s$ and $p$ electrons due to 
hybridization with $d$ electrons.
%
%
\begin{figure}[h]
\begin{center}
\includegraphics[width=12cm]{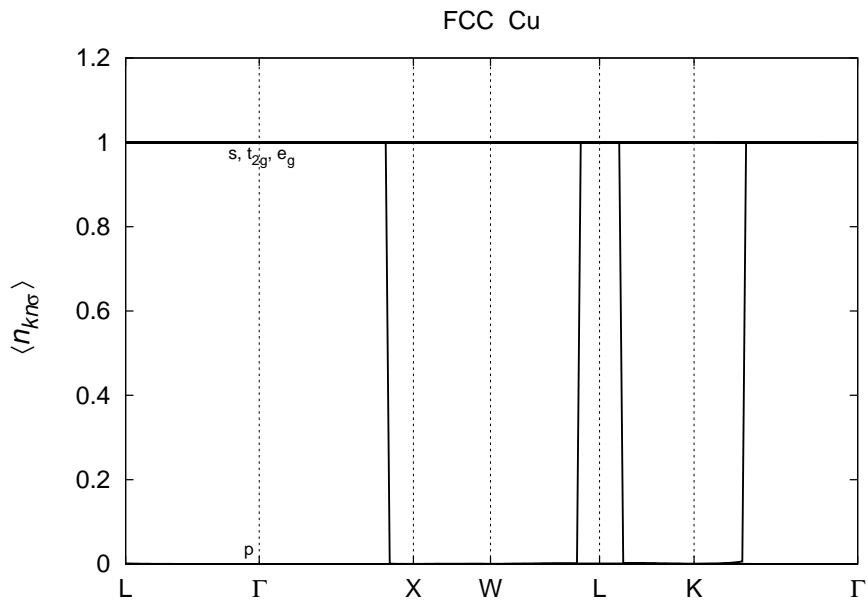}
\end{center}
\caption{Momentum distribution functions 
$\langle n_{kn\sigma} \rangle$ along high-symmetry lines for fcc Cu.  
}
\label{fignkncu}
\end{figure}
%
%

Finally we present in Fig. \ref{fignkncu} the MDF for Cu along 
high-symmetry lines.  
The $d$ bands of Cu are located far below the Fermi level, 
thus the MDF follow the FDF except a tiny deviation of the
branch for hybridized $pd$ electrons near 
$\boldsymbol{k}_{\rm F} = (0, 0.52, 0.52)$ on the K-$\Gamma$ line.  
The MDF of the conduction bands in Cu are described well by the band 
theory.
%
%
\begin{figure}[h]
\begin{center}
\includegraphics[width=12cm]{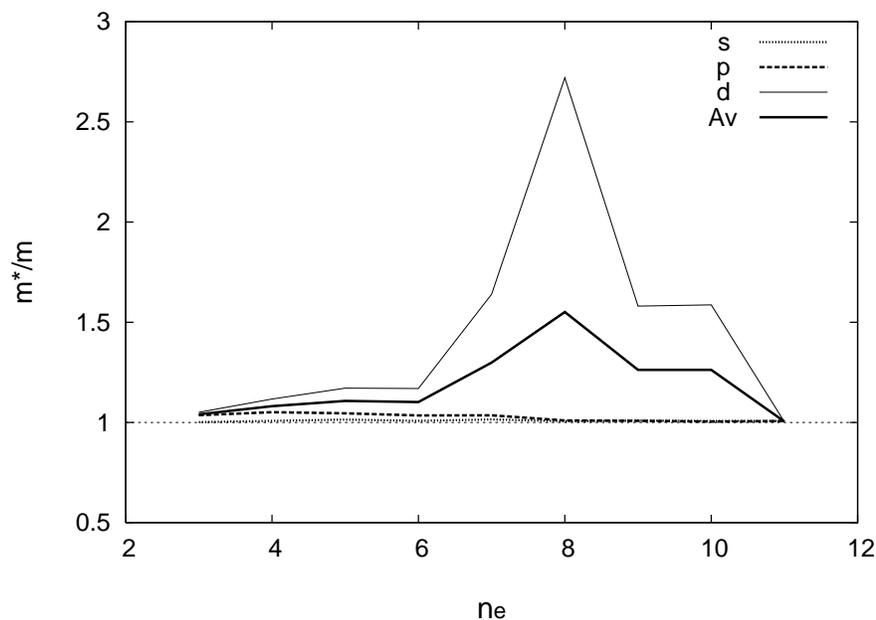}
\end{center}
\caption{Partial mass enhancement factors  
$m^{\ast}_{l}/m$ ($l=s$, $p$, and $d$) from Sc to Cu as a function of 
conduction electron number $n_{e}$.
Dotted curve: the MDF for $l=s$, dashed curve: the MDF for $l=p$, the
 thin solid curve: the MDF for $l=d$, solid curve: the average MDF 
$m^{\ast}/m$.  
}
\label{figmeffspd}
\end{figure}
%
%

\subsection{Mass enhancement factors}

The jump of the MDF on the Fermi surface provides us with the
quasiparticle weight, thus the mass enhancement factor.
We summarize in Fig. \ref{figmeffspd} systematic change of calculated  
partial mass enhancement factors (MEF) $m^{\ast}_{l}/m$ ($l=s$, $p$, 
and $d$). 
The deviations of the partial MEF for $s$ electrons 
from the Hartree-Fock value ($=1$) 
are only less than 1.5 \% from Sc to Cu; the $s$ electrons do not 
cause the mass enhancement.
The partial MEF for $p$ electrons 
are also close to one, though we find $3 \sim 5$ \% deviation 
from 1 for the elements from Sc to
Mn due to hybridization with $d$ electrons.
The partial MEF for $d$ electrons show a significant deviation
from the Hartree-Fock value except Sc and Cu in which the $d$ bands are
located above or below the Fermi level $\epsilon_{\rm F}$; 
$m^{\ast}_{d}/m = 1.117$ (fcc Ti),
$1.171$ (bcc V), $1.170$ (bcc Cr), $1.640$ (fcc Mn), $2.720$ (bcc Fe),
$1.581$ (fcc Co), $1.587$ (fcc Ni).  In particular, $m^{\ast}_{d}/m$ for
bcc Fe shows the maximum value 2.720 because the narrow $e_{g}$ bands
are located on the Fermi level $\epsilon_{\rm F}$.  
The $d$ electron contribution
therefore determines the systematic change of the average mass
enhancement $m^{\ast}/m$ via the relation 
$m/m^{\ast} = D^{-1} \sum_{l} (2l+1) m/m^{\ast}_{l}$ 
(see Eq. (\ref{eqnzlsum})).
We summarize in Table \ref{tmeff} the average MEF from Sc to Cu.  A
large value of bcc Fe is caused by the narrow $e_{g}$ bands on
$\epsilon_{\rm F}$.  We also calculated the MEF for
fcc Fe: $m^{\ast}/m=1.349$.  It is smaller than the bcc case since
there are no clear flat $d$ bands on $\epsilon_{\rm F}$.
%
%
\begin{figure}[h]
\begin{center}
\includegraphics[width=12cm]{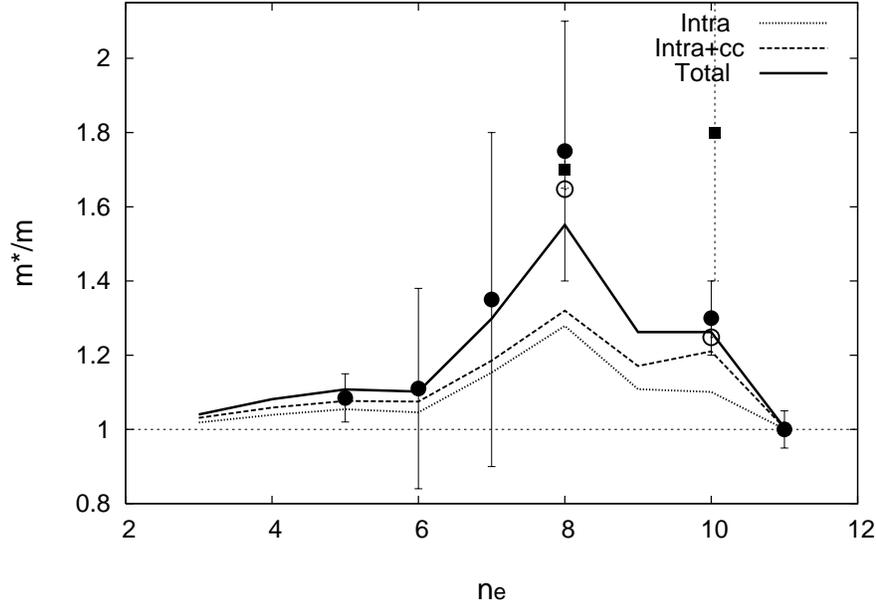}
\end{center}
\caption{Calculated mass enhancement factors ($m^{\ast}/m$) 
with/without three type of
 correlations as a function of conduction electron number 
$n_{e}$.  Dotted line: Intraorbital contribution, dashed line:
 intraatomic and interorbital charge-charge contributions, solid line:
 total mass enhancement including interorbital spin-spin contribution.
The total $m^{\ast}/m$ calculated with use of Anisimov's $U$ and
 $J$~\cite{anis97-2} are also expressed by open circles 
for Fe ($n_{e}=8$) and Ni ($n_{e}=10$).  
Experimental results obtained from the low-temperature specific heats
 (ARPES) data are shown by closed circles 
(closed squares)~\cite{papacon15,knapp72,beck70,cheng60,zimm61, 
weiss58,cho96,chio03,pepp01,jarl02,higashi05}. 
}
\label{figmecmp}
\end{figure}
%
%
%
%
\begin{table}[tb]
\caption{Calculated average mass enhancement factors in iron-group 
transition metals.
The results with parentheses for Fe and Ni are obtained with use of 
$(U, J) = (0.1691, 0.0662)$ Ry and $(0.2205, 0.0662)$ 
Ry~\cite{anis97-2}, respectively.
\vspace{5mm} }
\label{tmeff}
\begin{tabular}{cccccccccc}
\hline
Element  & Sc & Ti & V & Cr & Mn & Fe & Co & Ni & Cu \\ \hline
$m^{\ast}/m$ & 1.040 & 1.081 & 1.108 & 1.102 & 1.299 & 1.551 (1.648) &
			     1.262 & 1.262 (1.248) & 1.004 \\ 
\hline 
\end{tabular}
\end{table}
%
%

We examined the origin of the mass enhancement by considering three
types of correlation contributions.
Figure \ref{figmecmp} shows the result of analysis.  
Here we defined the intra-orbital contribution $m^{\ast}/m$(intra) by
$m^{\ast}/m$ when $\tilde{\zeta}_{LL^{\prime}} =
\tilde{\xi}_{lLL^{\prime}} = \tilde{\xi}_{tLL^{\prime}} = 0$, 
the inter-orbital charge-charge contribution $m^{\ast}/m$(intra+cc) 
by $m^{\ast}/m$ when   
$\tilde{\xi}_{lLL^{\prime}} = \tilde{\xi}_{tLL^{\prime}} = 0$, 
and the full value $m^{\ast}/m$(total) including the inter-orbital
spin-spin contribution.
The intra-orbital correlations make contribution 
to $m^{\ast}/m$(total) by about 50 \% irrespective of elements.  The
inter-orbital charge-charge correlations make a minor contribution for
the elements from Ti ($n_{e}=4$) to Fe ($n_{e}=8$) (see 
$m^{\ast}/m({\rm intra+cc}) - m^{\ast}/m({\rm intra})$).  
In Ni ($n_{e}=10$), the charge-charge contribution is enhanced, 
and becomes comparable to the intra-orbital contribution.
The inter-orbital spin-spin contribution 
(,{\it i.e.}, $m^{\ast}/m({\rm total}) - m^{\ast}/m({\rm intra+cc})$)
is comparable to the charge-charge contribution for 
the elements from Sc to Cr. 
It becomes significant for Mn, Fe, and Co.
These results indicate that the mass enhancements of Mn and Fe are
determined by the spin fluctuations (,{\it i.e.}, the intra-orbital 
plus inter-orbital spin-spin correlations), 
while the charge fluctuations 
(,{\it i.e.}, the intra-orbital plus inter-orbital charge-charge 
correlations) are important in the case of Ni.

The mass enhancement factor (MEF) of Fe has recently been investigated 
from the theoretical point of view.  
S\'{a}nchez-Barriga {\it et al.}~\cite{sanchez09} 
performed the three-body theory + LDA-DMFT calculations 
with use of $U=1.5$ eV and $J=0.9$ eV, and   
obtained $m^{\ast}/m = 1.25$ on the $\Gamma$-N line, which is  
too small as compared with the angle resolved 
photoemission spectroscopy 
(ARPES) result~\cite{sanchez09} $m^{\ast}/m = 1.7$.
Katanin {\it et al.}~\cite{katanin10} performed the finite-temperature 
LDA+DMFT calculations with use of the quantum Monte-Carlo technique 
(QMC) at 1000 K.  They obtained $m^{\ast}_{t2g}/m = 1.163$ for $t_{2g}$
electrons being in agreement with our result 
$m^{\ast}_{t2g}/m = 1.22$ (see Table \ref{tbccfemefft2g}).  
But the value for $e_{g}$ electrons 
was not obtained because of the non-Fermi liquid behavior due to 
strong spin fluctuations at finite temperatures, though we obtained
$m^{\ast}_{eg}/m = 1.67$ (see Table \ref{tbccfemeffeg}).
More recently, Pourovski {\it et al.}~\cite{pour14} 
reported the LDA+DMFT calculations of bcc Fe at 300 K using the 
continuous-time QMC technique.  
They obtained the average mass enhancement 
$m^{\ast}/m \approx 1.577$ being in good agreement with the 
present result $m^{\ast}/m = 1.551$.   

We note that the first-principles Gutzwiller theory underestimates 
the mass enhancement factor.  The LDA+Gutzwiller calculations by 
Deng {\it et al.}~\cite{deng09} yield a reasonable value
$m^{\ast}/m \approx 1.564$, but they adopted too large a Coulomb
interaction parameter $U=7.0$ eV.  Recent calculations 
based on the LDA+Gutzwiller theory with reasonable values 
$U=2.5$ eV and $J=1.2$ eV 
result in $m^{\ast}_{eg}/m \approx 1.08$ for $e_{g}$ electrons and 
$m^{\ast}_{t2g}/m \approx 1.05$ for $t_{2g}$ electrons~\cite{borghi09}. 
These values are too small as compared with the present results 
$m^{\ast}_{eg}/m = 1.67$ and $m^{\ast}_{t2g}/m = 1.22$ 
and too small as compared with 
the ARPES value~\cite{sanchez09} $m^{\ast}/m = 1.7$.

Experimental data obtained by the $T$-linear electronic specific heat
and the ARPES are also shown in Fig. \ref{figmecmp}.
Experimentally, Sc shows the hcp structure.  The MEF of
hcp Sc~\cite{papacon15} 
estimated from the low-temperature specific heat and the density
of states at $\epsilon_{\rm F}$ is 2.04.
This includes the MEF due to electron-phonon interaction, 
$1+\lambda_{ep}$, where $\lambda_{ep}$ denotes the electron-phonon
coupling constant.  A simple way to remove the effect is to measure the
$T$-linear electronic specific heat above the temperatures larger than
the Debye temperature $\Theta_{\rm D}$.  Then we obtain the experimental
value due to electron correlations~\cite{knapp72} 
$m^{\ast}_{\rm expt}/m = 1.44$.  
The present result for the fcc Sc is $m^{\ast}/m = 1.040$, 
and is smaller than the experimental value 1.44 for the hcp Sc.
The titanium also shows the hcp structure.  Taking the same step, we
find the electronic contribution of the MEF~\cite{beck70}, 
$m^{\ast}_{\rm expt}/m = 1.19$, which is considerably larger than the
present result $m^{\ast}/m = 1.081$ for fcc Ti.

The vanadium shows the bcc structure, so that we can directly compare
the present result with the experimental one.  The calculated result
$m^{\ast}/m = 1.108$ is consistent with 
$m^{\ast}_{\rm expt}/m = 1.02 \sim 1.15$ in which the MEF due to
$\lambda_{ep}$ has been eliminated~\cite{beck70}.
The MEF of bcc Cr are estimated from the low-temperature specific heat
data and DOS at $\epsilon_{\rm F}$~\cite{papacon15,cheng60}; 
$m^{\ast}_{\rm expt}/m = 0.84 \sim 1.38$.
The MEF due to the electron-phonon interaction are not eliminated there.
The calculated result $m^{\ast}/m = 1.102$ is in the range of the
experimental values.

The MEF of fcc Mn can be estimated by an extrapolation of the specific
heat data for Mn-Cu alloys~\cite{zimm61,papacon15}; 
$m^{\ast}_{\rm expt}/m = 0.9 \sim 1.8$.
The experimental value $m^{\ast}_{\rm expt}/m = 1.4$ obtained from the
high-temperature specific heat of $\gamma$-Mn~\cite{weiss58} 
is also in this range.
The present result $m^{\ast}/m = 1.299$ does not contradict with these
data. 

The MEF of the bcc Fe estimated from the low-temperature specific
heats~\cite{papacon15,cheng60,cho96,chio03,pepp01} 
are $m^{\ast}_{\rm expt}/m = 1.4 \sim 2.1$.
The present result $m^{\ast}/m = 1.551$ and the result 
$m^{\ast}/m = 1.648$ obtained with use of Anisimov's $U$ and
$J$~\cite{anis97-2} are  
consistent with the experimental data.  
The result is also consistent with the experimental
value  $m^{\ast}_{\rm expt}/m = 1.7$ obtained by 
ARPES~\cite{sanchez09}.

The cobalt shows the hcp structure below 700 K.  
The MEF of hcp Co estimated from the low
temperature specific heat~\cite{papacon15} is 2.3.  
Using the MEF of electron-phonon
coupling~\cite{jarl02} 
$1.2 \sim 1.4$, we find the experimental value 
$m^{\ast}_{\rm expt}/m = 1.6 \sim 1.9$.  The present result for fcc Co 
$m^{\ast}/m = 1.262$ is smaller than the hcp experimental 
value $1.6 \sim 1.9$.

The experimental MEF of fcc Ni estimated from the low-temperature 
specific heat and
the DOS at $\epsilon_{\rm F}$~\cite{papacon15} is 1.7.  
When we remove the electron-phonon MEF 
$1+\lambda_{ep} \sim 1.3$ estimated from the ARPES~\cite{higashi05}, 
we find $m^{\ast}_{\rm expt}/m \approx 1.3$.  
The present result $m^{\ast}/m = 1.262$
and the result $m^{\ast}/m = 1.248$ obtained with use of Anisimov's $U$
and $J$~\cite{anis97-2} 
are in agreement with the experimental value 1.3, but is smaller 
than the values 
$1.4 \sim 2.2$ obtained by the ARPES~\cite{higashi05}.

Although the quantitative comparison between the theory and experiments
is not easy at the present stage, the present results seem to be
consistent with the experimental data.
The underestimate of the MEF in Ni in comparison with ARPES data 
may be attributed to the
magnon mass enhancement~\cite{fulde83, fulde12,hofmann09} 
which is not taken into account in the present theory.

\section{Summary and discussions}

We have investigated the momentum distribution function (MDF) of
iron-group transition metals from Sc to Cu on the basis of the
first-principles momentum-dependent local ansatz (MLA) wavefunction 
method, which we recently developed for quantitative calculations of the
ground-state properties. 

The MDF in the real system depends on the
momentum $\boldsymbol{k}$ via both the eigenvectors
$u_{Ln\sigma}(\boldsymbol{k})$ and the energy eigenvalue 
$\tilde{\epsilon}_{kn\sigma}$ measured from the Fermi level.
We obtained the MDF along high-symmetry lines of the first Brillouin
zone, and analyzed them with use of the partial MDF.
In iron-group transition metals, $3d$ correlated electrons play an 
important role in the MDF.  The average Coulomb (exchange)
interaction $U$ ($J$) increases linearly from 0.1 (0.04) Ry to 0.3
(0.07) Ry with increasing conduction electron number $n_{e}$ from Sc to
Cu, and the $d$ electron band width gradually decreases from 0.4 Ry to
0.3 Ry.  Thus $d$ electron correlations become important 
with increasing $n_{e}$. 
The correlation effects on the MDF however occur via the $d$ 
electrons near the Fermi surface.  The MDF for Cu therefore 
behave as an independent electron system because the $d$ bands are
located far below the Fermi level. 

We verified that the MDF for Sc 
follow approximately the Fermi-Dirac distribution function (FDF) 
for independent electrons.
In Ti, V, and Cr, we found small deviations of the MDF from the 
FDF along high-symmetry lines.
From Mn to Ni, there exist significant deviations of the MDF from 
the FDF due to electron correlations.
In these systems, the MDF for $d$ electrons show a strong momentum 
dependence along high-symmetry lines, while those for $sp$ electrons 
show small deviations from the FDF via hybridization between $sp$ 
and $d$ electrons. 

We found that bcc Fe shows the largest deviation of the MDF from the
FDF because the narrow $e_{g}$ bands with flat dispersion are located on
the Fermi level.  Accordingly, the $d$-electron partial MDF for bcc 
Fe shows the
strong momentum dependence via the energy $\epsilon_{k}$.
We verified that the MDF for fcc Fe shows less deviation from the FDF 
since there is no such a narrow band on the Fermi level.  
In the case of Ni, we found that a large deviation of the MDF with 
$t_{2g}$ symmetry appears along the X-W line because of the 
existence of the flat $t_{2g}$ energy bands on the Fermi level 
along the line and strong
electron correlations.

We obtained the momentum-dependent mass enhancement factors (MEF)
from the jump of the MDF at the Fermi surface.
Calculated average MEF show considerably large enhancement from Mn 
to Ni more than 1.2, while the other metals show small enhancement 
less than 1.2.
The results seem to be consistent with the experimental values, though 
there is an ambiguity in estimating the electronic 
contributions of MEF from the experimental data.  
We found that bcc Fe shows the largest MEF $1.55 \sim 1.65$ due to
$e_{g}$ electrons with narrow bands on the Fermi level.  
Calculated MEF for bcc Fe is in good agreement 
with the recent results based on the LDA+DMFT at finite temperatures 
as well as the experimental result obtained by ARPES.
The MEF in fcc Fe is found to be 1.35 which is smaller than the bcc 
value. 
We found that the mass enhancements for Mn, Fe, and Co are mainly 
caused by spin
fluctuations, while the mass enhancement for Ni is caused by charge
fluctuations.  For the other metals, both spin and charge fluctuations
contribute to the small mass enhancements.

In the present calculations, we assumed the paramagnetic state from Sc
to Cu, though the transition metals from Cr to Ni show the magnetic 
order at the ground state.
The bcc Fe, for example, shows the ferromagnetism.  
In the paramagnetic state, the $t_{2g}$ electron bands are located below
the $e_{g}$ electron bands by 0.12 Ry. 
When the bcc Fe is
spin polarized, we expect that the weight of $t_{2g}$ electrons with
smaller $m^{\ast}_{kn\sigma}/m$ is increased on the Fermi surface as
compared with the $e_{g}$ electrons due to exchange splitting.
Thus the average MEF is expected to be reduced by $5 \sim 10$ \% due 
to spin polarization.

The second point which we have to remark is that the present theory is
based on the single-site approximation (SSA); it does not take into
account the nonlocal correlations.  Long-range spin fluctuations 
are known to cause additional magnon mass enhancement~\cite{fulde12}, 
which can cause the logarithmic divergence in the vicinity of 
magnetic instability point. 

Direct observation of the MDF by means of the energy integration of
ARPES data is highly desired in order to verify the quantitative 
agreement between the theory and experiment.
Calculations of the MDF and MEF for Fe, Co, and Ni in the 
ferromagnetic state and the development of 
the theory to the nonlocal case are left for future work
towards quantitative understanding of the ground-state properties of
iron-group transition metals.

\begin{acknowledgment}


The authors would like to express their sincere thanks to Prof. P. Fulde
 for his valuable comments and encouragements to the present work. 
This work is supported by a Grant-in-Aid for Scientific Research
 (25400404). 
Numerical calculations have been partly carried out with use of the 
facilities of the Supercomputer Center, the Institute for Solid State 
Physics, the University of Tokyo.

\end{acknowledgment}



\end{document}